\documentclass[sigconf]{acmart}
\AtBeginDocument{%
  }

\setcopyright{acmlicensed}
\copyrightyear{2024}
\acmYear{2024}
\acmDOI{XXXXXXX.XXXXXXX}
\acmConference[Conference '24]{Conference}{December 12--13,
  2024}{Louvain-la-Neuve, Belgium}
\acmISBN{978-1-4503-XXXX-X/18/06}

\usepackage{pifont}
\usepackage{accents}
\usepackage{natbib}
\usepackage{subfigure}
\usepackage{xspace}
\newcommand{\ie}{i.e.,\xspace}
\newcommand{\eg}{e.g.,\xspace}
\newcommand{\vs}{vs.\xspace}

\newcommand{\tp}{\textsc{Taoist}\xspace} 
\begin{document}

\title{A Model-based Approach to Assess Regular, Constant, and Progressive User Interface Adaptivity}

\author{Alaa Eddine Anis Sahraoui}
\email{alaa.saharaoui@uclouvain.be}
\orcid{0009-0001-5402-9605}
\affiliation{%
  \institution{Universit\'e catholique de Louvain}
  \department{Louvain Research Institute in Management and Organizations (LouRIM)}
  \streetaddress{Place des Doyens, 1}
  \city{Louvain-la-Neuve}
  \postcode{B-1348}
  \country{Belgium}}

\renewcommand{\shortauthors}{Alaa Eddine Anis Sahraoui}

\begin{abstract}
  Adaptive user interfaces adapt their contents, presentation, or behavior mostly in a sudden, fluctuating, and abrupt way, which may cause negative effects on the end users, such as cognitive disruption. Instead, adaptivity should be regular, constant, and progressive. To assess these requirements, we present \tp, a hidden Markov model-based approach and software environment that seek the longest repeating action subsequences in a task model. The interaction state space is discretely produced from a task model and the interaction observations are dynamically generated from a categorical distribution exploiting the subsequences. Parameters governing adaptivity and its results are centralized to support two scenarios: 
intra-session for the same user and inter-session for the same or any other user, even new ones. The end-user can control the adaptivity when initiated by accepting, declining, modifying, postponing, or reinitiating the process before propagating it to the next iteration. We describe the \tp implementation and its algorithm for adaptivity. We illustrate its application with examples, including the W3C reference case study. We report the results of an experiment that evaluated \tp with a representative group of ten practitioners who assessed the regular, constant, and progressive character of adaptivity after four intra-session iterations of the same task.
\end{abstract}


\begin{teaserfigure}
\centering
\includegraphics[width=\textwidth]{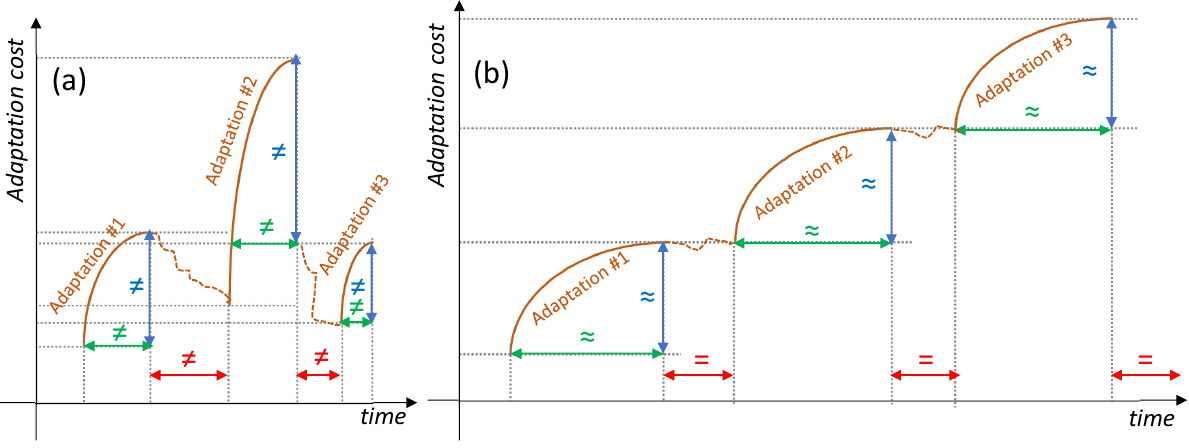}
\vspace{-16pt}
\caption{User Interface adaptation: (a) without \tp, it is sudden ($\neq$ in red), fluctuating ($\neq$ in  blue), and abrupt ($\neq$ in green); 
(b) with \tp, it is regular ($\approx$ in red), constant ($\approx$ in blue), and progressive ($\approx$ in green).}
\label{fig:requirements}
\vspace{+8pt}
\end{teaserfigure}

\received{12 December 2024}

\maketitle

\section{Introduction}
Adapting the Graphical User Interface (GUI) of an interactive application 
to a particular context of use~\cite{Calvary:2003} has been largely investigated in the literature~\cite{Eisenstein:2000,Repo:2004,Shankar:2007,Soui:2015,Yigitbas:2016,Hussain:2018,Zouhaier:2021} and applied in many domains~\cite{Repo:2004,Keranen:2002,Jones:2016}. 
Yet, this problem cannot be considered solved~\cite{Alvarez:2009,Jiang:2022} 
despite the tremendous progress made since the early days~\cite{Browne:1990,Dieterich:1994,Oppermann:1997}. 
The adaptation should always be for the ultimate benefit of the end-user,
such as to improve the performance~\cite{Gajos:2006,Reguera:2021}, to address their preferences~\cite{Eslami:2018}, and habits~\cite{Gajos:2008}. 
Despite this laudable goal, adaptation faces various problems: 
(1) the expected benefit is not always forthcoming and is small compared to the overall cost of adaptation~\cite{Lavie:2010}; 
(2) any adaptation engenders a cognitive disruption~\cite{Hui:2009}; 
(3) adaptation is effective only under certain conditions and is highly dependent on intermingled factors, such as personal traits~\cite{Gajos:2017}; 
(4) the existing interaction history is rarely exploited for a current user or new users, who do not have such a history~\cite{Alvarez:2009}.

We believe that adaptation suffers from three main limitations that arise from sudden, fluctuating, and abrupt changes (\autoref{fig:requirements}-a). Firstly, adaptation is \textit{sudden} because it appears at different moments in time and at variable locations in the interface, making it hardly predictable \cite{Gajos:2008}. Secondly, it is \textit{fluctuating} because its intensity is variable on a small part or a large portion of the interface, thus endangering consistency~\cite{Biermeier:2021}. Thirdly, it is \textit{abrupt} because it induces a significant change in the interface~\cite{Hui:2009}. Rather than working only on taking into account user-related parameters and introducing even more adaptation rules that are challenging to discover~\cite{Nasfi:2014}, we argue that adaptation should be \textit{regular} instead of sudden, \textit{constant} instead of fluctuating, and \textit{progressive} instead of abrupt (\autoref{fig:requirements}-b).


We believe that the human being, who benefits from a real ability to adapt habits depending on the context of use, does not always want to adapt, but is required to do so, thereby preferring that adaptation should satisfy three requirements among others~\cite{Ghannem:2017,Alvarez:2009}:
\begin{enumerate}
    \item \textit{Adaptation should be regular}: the adaptation process should follow a recognizable pattern. In GUIs, we interpret that adaptation should occur at evenly spaced times, GUI locations, like a regularly scheduled meal for health.
    It should not adapt the entire UI at once, but propose a series of reasonably scaled adaptation steps that form a recognizable pattern.
    A positive example of a regular adaptation is \href{https://www.youtube.com/watch?v=P5jhgl5EQ3A}{\textsc{MiniAba}}~\cite{Schlee:2004}, where the end user designates similar portions of the UI to include or exclude for each adaptation, thus posing a regular and recognizable action.
    
    \item \textit{Adaptation should be constant}: the adaptation process should occur continuously with a similar intensity over time. 
    In GUIs, we interpret that adaptation should keep the same level of change over time, preferably limited to minimize any disruption.
    A positive example of a constant adaptation is \textsc{WiSel}~\cite{Mezhoudi:2015}, where the adaptation consists of a continuous suite of widget adaptations (\eg by selection, by substitution, by progressive enhancement, or by graceful degradation) converging to a solution where no adaptation is required anymore.  
    
    \item \textit{Adaptation should be progressive}: the adaptation process should gradually move from the current state (before adaptation) to a future state (after adaptation) following several steps instead of a single step. In GUIs, we interpret the adaptation to consist of a series of gradually evolving steps.
    A positive example is \href{https://userinterfaces.aalto.fi/adaptive/}{progressive adaptivity}~\cite{Todi:2021} of a graphical pull-down menu decomposed into several successive steps that are enacted one after another rather than all at once.  
\end{enumerate}

To assess these requirements, 
we present \tp (\underline{T}ask-based \underline{A}daptation \underline{o}f Graph\underline{i}cal U\underline{s}er In\underline{t}erfaces), 
a model-based approach and software environment for GUI adaptation at run-time (adaptivity) combining two contributions: (1)~a task model following W3C notation~\cite{W3C-Task:2014} that derives potential Abstract User Interfaces (AUIs)~\cite{W3C-Abstract:2014}, to be directly and partially instantiated as Final User Interfaces (FUIs); (2)~a Machine Learning component consisting of a Hidden Markov Model that represents the interaction sequences generated from the task model to identify in these sequences the Longest Repeating action Subsequences collected from end users to propagate them from one session to another.

The remainder of this paper is organized as follows:
\autoref{sec:limitations} reviews the state-of-the-art in adaptation approaches in Model-Based User Interface Design (MBUID) and ML, and discusses some limitations that motivate this work (\autoref{sec:motivations}).
\autoref{sec:taoist} defines our theoretical background, provides an overview of the adaptation process, and describes the two main components of \tp, the software that supports this approach.
\autoref{sec:case-studies} illustrates this approach on case studies, such as the W3C reference case. 
\autoref{sec:performance} evaluates the performance of \tp in terms of execution time, number of nodes, and number of adaptation solutions.
\autoref{sec:qualitative} reports a qualitative study evaluating how end users perceived \tp adaptations. 
Finally, \autoref{sec:conclusion} summarizes this work and presents some ideas for future work.

\section{Related Work} 
\label{sec:limitations}
On one hand, \textit{model-based approaches} ~\cite{Eisenstein:2000,Martin:2017,Hussain:2018,Josifovska:2019} and \textit{model-driven engineering}~\cite{Akiki:2014,Akiki:2016,Yigitbas:2020,Bouraoui:2019} for GUI adaptation~\cite{Bogdan:2022} adopt a \textit{top-down approach}, where ``white box'' models are created to reason about adaptation and its components~\cite{Yigitbas:2019:component}. These models seek to estimate~\cite{Gajos:2006}, predict~\cite{Gajos:2008}, derive~\cite{Welivita:2016}, and evaluate~\cite{Soui:2015,Yigitbas:2019:feedback} adaptations. Rule-based~\cite{Nasfi:2014,Sousa:2008} and knowledge-based \cite{Soui:2017,Puerta:1997} systems define the rules to be selected, parsed, and executed to adapt a GUI to its context of use. Consequently, the number of rules and the volume of knowledge required to adapt increase with time and complexity. Earlier techniques needed project recompilation~\cite{Schlee:2004}, thus increasing cost, delay, and risk~\cite{Lavie:2010,Asthana:2013}.

For example, \citet{Blouin:2011} combined aspect-oriented modeling with property-based reasoning to reduce the design space of possible adaptations and provide an efficient way to adapt a GUI depending on its contest of use and available GUI components.
Facing the same explosion of possible GUI adaptations, \citet{Duhoux:2019} relied on feature-based context-oriented programming to model the context of use and its variations, as well as the adaptations required by contextual variations. each being captured in a separate feature model. This family of feature models is richer than using a single feature model as in \textsc{MiniAba}~\cite{Schlee:2004}, which is limited to a feature model regulating the inclusion or exclusion of GUI elements depending on contextual conditions.
 
\citet{Peissner:2012} developed the \textsc{MyUI} framework to increase accessibility through the generation of adaptive UIs. The framework performs run-time adaptations to diverse user needs and device usage by personalizing GUI presentation, layout, modalities, as well as navigation path. It uses a multimodal design pattern repository that serves as the basis for creating personalized UIs. For creating adaptive UIs, \textsc{MyUI} is based on adaptation rules defined by design patterns, and their combinations are manually handled during conception.

\citet{Hussain:2018} presented a model-based approach that is designed to generate adaptive GUIs. Their system is implemented as an Adaptive User Interface User Experience Authoring (A-UI/UX-A) tool. This tool is based on ontology models that allow the representation of the user, context, and device. Moreover, the A-UI/UX-A tool incorporates a component for authoring adaptation rules. The construction of the A-UI/UXA tool requires two processes: an offline process to create models and generate basic UI adaptation rules; and, an on-line process to allow generating adaptive GUIs.

\citet{Bouraoui:2019} also introduced a model-based approach that adapts GUIs for multi-platforms, multi-devices, and hybrid cross-platforms. It focuses on the semi-automatic generation of accessible UIs based on the MDD framework. To allow UI generations, their approach relies on some transformation techniques borrowed from model-driven engineering. These techniques enable the transformation from abstract accessible UI models to executable GUIs that meet specific accessibility requirements. Classic tools including Ecore and Eclipse Modeling Framework (EMF) are considered for metamodeling and for creating the source models and the needed transformations, respectively. 

\citet{Yigitbas:2020} experimented a model-driven approach for self-adaptive UIs based on the latest context of use. The authors introduce two Domain Specific Languages (DSL), including ContextML and AdaptML for defining context properties and for modeling adaptation rules, respectively. Context information is extracted at run-time and used for selecting adaptation rules. To allow for the adaptation of the GUI in runtime, the approach relies on several models: a context model to generate context services, an adaptation model for generating adaptation services, and an abstract UI \cite{W3C-Abstract:2014} and domain models for generating the final UI. \citet{Tran:2012} showed that the spectrum of such abstract UIs can be very large and that finding the best adaptation in this spectrum is an elusive problem as this represents an overconstrained multi-objective optimization problem.

On the other hand, ML approaches~\cite{Langley:1997,Welivita:2016,Lee:2011,Mezhoudi:2021,Silva:2021,Zouhaier:2021,Jiang:2022} mainly adopt a \textit{bottom-up approach}, where data are acquired from the context of use to feed a ``black box'' model able to derive potential adaptations. ML approaches expect to produce a model that fits the context of use from its actual description, not the other way around, as in model-based approaches. Unlike the latter, ML faces the problem of an implicit \vs explicit model, which does not facilitate reasoning about the model beyond classification and prediction.

In these approaches, adaptation is mostly investigated as a one-shot process (\ie an adaptation is performed once, either at design time or at run time) or as a few-shot process (\ie an adaptation is performed only a few times at dedicated moments, such as when an interactive session starts). Consequently, the cost of adaptation could be high at the price of some shortcomings for the end user, leading to a suboptimal cost/benefit ratio~\cite{Lavie:2010}. A particular model improves user performance~\cite{Reguera:2021} at the price of paying a disruption in the user's mental model~\cite{Hui:2009}. Any adaptation will always have the effect of disturbing the end-user to some extent - this is unavoidable - but by introducing the adaptation gradually, the disturbance could be minimized by distributing it over several iterations. 

For example, Reinforcement Learning (RL) of Adaptive User Interfaces for Accessibility Context \cite{Zouhaier:2021} exploits three Knowledge Layers (KL)s depending on the target adaptation: the Disability Knowledge Layer learns the behavior of the structure of the UI depending on the disability profile, the Modality Knowledge Layer learns adaptation facilities based on the couple (UI, profile), and the Platform Knowledge Layer exploits platform knowledge to learn adaptation facilities.

RL can also train agents to learn how to adapt UIs in a specific context of use to maximize the user engagement by using an interaction model in a reward function \cite{Gaspar:2024}. \textsc{Marlui} \cite{Langerak:2024} engaged a user agent that mimics a real user and learns to interact via point-and-click actions with a UI agent that learns interface adaptations, to maximize efficiency by observing the user agent's behavior.

RL seems particularly well-suited to adapting the interface, no longer predictively on the basis of predefined models, but statistically on the basis of models obtained at runtime, which takes account of the user's real and not just modeled behavior. However, nothing is said about the extent to which this adaptation needs to be regulated in time, space, and effort.

More recently, adaptation has been understood as a continuous process of small discrete steps~\cite{Todi:2021} rather than a discrete process of a single large scheme. In this way, adaptation is continuously supported to detect when adaptation is needed and decide which GUI part adaptation should be applied~\cite{Abrahao:2021}. In this way,  the adaptation cost is distributed into a sequence of small costs distributed in time and space throughout the process.

The literature defines various adaptation processes, ranging from adaptability~\cite{Oppermann:1997} initiated by users, to the adaptivity executed by the system~\cite{Akiki:2014}, implicit~\cite{Leiva:2011} instead of explicit, along with hybrid adaptations that mix both initiatives~\cite{Horvitz:1999}. The advantages and drawbacks of each adaptation process are widely discussed in the literature \cite{Oppermann:1997,Calvary:2003,Hui:2009,Lavie:2010,Asthana:2013,Mezhoudi:2021}. 
The lack of accurate and transparent adaptation decisions~\cite{Gajos:2006}, limited user controllability~\cite{Lavie:2010}, feedback~\cite{Mezhoudi:2015} and predictability~\cite{Gajos:2008} are considered the most serious concerns.

Therefore, in the current computational landscape ~\cite{Jiang:2022} of real-time adaptation and context awareness~\cite{Dubiel:2022,Motti:2013}, the support of GUI adaptation at run-time~\cite{Blumendorf:2010,Yigitbas:2019:fly} becomes the crucial requirement to handle varying resources, evolving user needs and system settings. An efficient implementation of adaptation that considers changing user preferences and takes several context variations at run-time is still challenging~\cite{Abrahao:2021,Leiva:2018}. Due to the complexity of modeling and predicting user preferences and profiles, continuous user involvement is a critical requirement to increase satisfaction and improve usability.  So far there are few works that effectively involve ML techniques in finding the appropriate adaptation and in satisfying the requirements. Some works do involve some ML techniques, such as Reinforcement Learning Model~\cite{Todi:2021,Gaspar:2023}, but they are still in their infancy.

\begin{figure*}
    \centering
    \includegraphics[width=\textwidth]{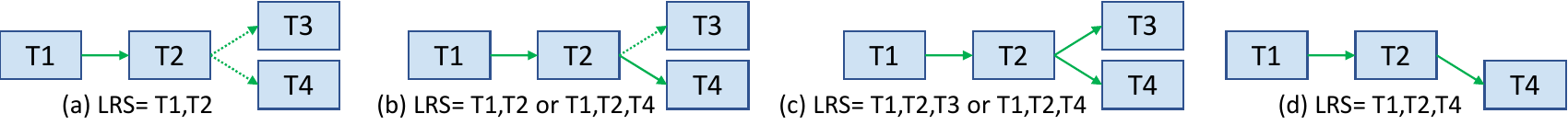}
    \caption{Examples of Longest Repeating Subsequences in a given GUI.}
    \label{fig:hmm-example}
\end{figure*}
\section{Motivations}
\label{sec:motivations}
User interface (UI) adaptation is a crucial aspect of modern software design~\cite{Akiki:2016,Wang:2024}, aiming to enhance user experience~\cite{Hussain:2018} (which can be measured through electroencephalograms~\cite{Gaspar:2023}, satisfaction \cite{Miraz:2021}, and its perception~\cite{Deuschel:2016}. Implementing UI adaptation in a regular, constant, and progressive manner is motivated by several reasons.

\textbf{The principle "Learn once, use many" should be preserved}: end users naturally develop familiarity with UIs, especially when consistency is preserved. Step-by-step adaptations ensure that these established habits are not disrupted suddenly.

\textbf{The interactive tasks should remain operationally stable}: gradual adaptations ensure that interactive tasks that are part of business processes are not disrupted by sudden UI shifts, maintaining productivity and operational stability.

\textbf{The user feedback should be incorporated in the loop}: progressive adaptations allow for continuous user feedback, enabling the UI to refine and improve itself based on real-world usage and preferences.

\textbf{The adaptation should iteratively bring improvements}: smaller adaptations are easier to test, to be evaluated by end-users and to be refined iteratively (\eg undoing a small adaptation is more reversible and less expensive than a large one), leading to a more usable version as adaptation evolves.

\textbf{Adaptation mistakes should be identified and corrected as soon as possible}: gradually adapting UI should make it easier to identify and correct adaptation errors, mistakes, mishaps, or issues specific to each adaptation stage, minimizing the risk of large-scale failures. If a change leads to unforeseen problems, its limited scope ensures that the impact on the overall software and user base is minimized. Moreover, large-scale adaptation failures will propagate more extensively than small ones and are more complex to undo.

\textbf{Adaptation should foster training}: end users can be educated about new features by adaptation in a manageable way, ensuring that they understand and can utilize them, therefore facilitating the transition from novice to expert users.

\textbf{Adaptation should maintain end-user's engagement}: gradual adaptations can keep end users engaged in their interaction with the software,  without feeling overwhelmed.


\section{GUI Adaptation by Markov Chains and Longest Repeating Sequences}
\label{sec:taoist}

\subsection{Theoretical Background}

Hidden Markov Chains or \textbf{\textit{Hidden Markov Models}} (HMMs) are sequence models characterized by a graph where nodes express a probability distribution \cite{Young:2013} over labels and edges give the probability of transitioning from one node to the other~\cite{Franzese:2019}. HMMs have been used extensively to study stochastic processes, making them suitable for studying the end-user's interaction with a GUI over time~\cite{Deshpande:2001} and its usability~\cite{Thimbleby:2001}. An HMM is characterized by a graph where nodes express a particular GUI state (\eg a web page, a form, a particular field) and edges capture the probability of navigating from a GUI state to another (\eg navigating from a web page, a window, a field to another). A sequence model~\cite{McComb:2017} can be obtained from log files to predict to which web page the end user is likely to navigate or from the GUI event list to predict to which window or field the end user is likely to move. The simplest HMM form is the \textit{first-order Markov HMM}~\cite{Franzese:2019}, which predicts the next user's action by examining the last performed action. The model order is equal to the number of actions: a $k^{\text{th}}$-order model considers the last $n$ actions performed by the user. High-order models are expected to give better predictions than first-order models, as the latter do not look far enough in the interaction history.

To support sequence learning~\cite{McComb:2017}, a \textbf{\textit{Longest Repeating Subsequence}} (LRS)~\cite{Pitkow:1999} is defined as the longest repeating sequence of end-user actions where the number of consecutive items repeats more than a given threshold $T$, which is typically unitary ($T{=}1$). We rely on this mechanism because it remains in compliance with HMMs while maintaining simplicity, understandability, and predictability~\cite{Kuchinski:2019}. \autoref{fig:hmm-example} shows a hypothetical GUI with four windows $T1$, $T2$, $T3$, and $T4$. If the end users go from window $T1$ to $T2$, one user will move from $T2$ to $T3$ and another from $T2$ to $T4$, then the LRS will be $LRS{=}T1,T2$ (\autoref{fig:hmm-example}-a). If more than one user moves to $T4$, then $LRS{=}T1,T2$ or $LRS{=}T1,T2,T4$ since $T4$ is predominant over $T3$ (\autoref{fig:hmm-example}-b). When both $T3$ and $T4$ were subject to interaction more than once, then $LRS{=}T1,T2,T3$ or $LRS{=}T1,T2,T4$ (\autoref{fig:hmm-example}-c). $T4$ is the only window finally used, so $LRS{=}T1,T2,T4$ (\autoref{fig:hmm-example}-d).

\subsection{Overview of the GUI Adaptation Process}
\label{sec:process}
Based on the two above components, our model-based approach to assess regular, constant, and progressive adaptivity consists of:
\begin{enumerate}
    \item \textit{Creating a task model}: a task model of the GUI of interest is created using the notation recommended by the W3C~\cite{W3C-Task:2014} and stored in its XML format~\cite{Limbourg:2004}.
    \item \textit{Creating a starting Markov model}: based on the configuration of the task model, a first-order Markov model is created corresponding to the first leftmost sub-task/action in the task model. 
    At this time, we usually specify the threshold for repeated items for LRS as $T{=}1$. This step initializes the LRS with the first node.
    \item \textit{Generating an Abstract User Interface}: using the task model, a model-to-model transformation generates an Abstract User Interface (AUIs) according to the W3C recommendation for AUIs~\cite{W3C-Abstract:2014}.
    \cite{Tran:2012} perform a systematic generation of all possible AUIs from which the designer has to choose one or many of them, among which the end user has to choose only one. Instead of using such an extensive and expensive approach, we focus on the dynamic generation at run-time of one AUI at a time, the one corresponding to the current subtask.

    \item \textit{Generating a Final User Interface}: based on the previous AUI, a \textit{fractional reification} instantiates the AUI into a fractional final UI (FUI) at runtime. Instead of transforming an AUI into a Concrete User Interface (CUI)~\cite{Calvary:2003} into a final one, we skip the CUI level by generating the GUI code corresponding to the current AUI.
\begin{figure*}
    \centering
    \includegraphics[width=.67\textwidth]{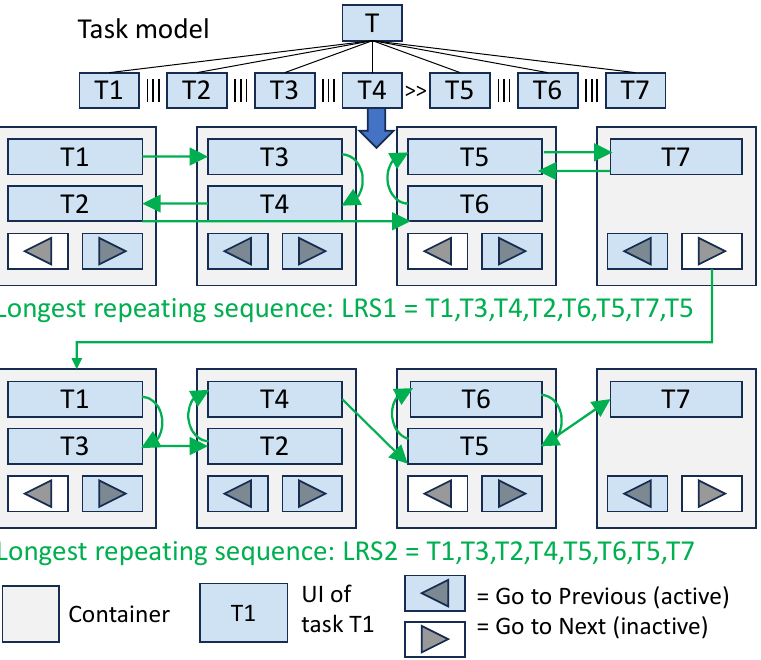}
    \vspace{-6pt}
    \caption{An example of GUI adaptation for the inter-session scenario.}
    \label{fig:adaptation-ex1}
\end{figure*}
\item \textit{Executing the Final User Interface}: the end user is then presented with the final UI generated at run-time. Depending on the temporal operators contained in the task model, controls are offered to navigate to another portion of the task model. 
    Depending on the end user's choice, a new subtask/action will be selected.
    \item \textit{Creating a $k^{\text{th}}$-order Markov model}: the task selection extends the initial first-order Markov model into a $k^{\text{th}}$-order Markov model, where $k$ evolves with the number of adaptations ($k{=}2,3,...$). The newly selected subtask/action then inserts a new edge in the graph with its corresponding transition as in \autoref{fig:hmm-example}. Only subtasks/actions that are subject to activation, navigation, and execution will feed the model and will be pruned in the LRS. In this way, we avoid the overwhelming consideration of all possible AUIs~\cite{Tran:2012}. 
    By pruning the initial data set while keeping only LRSs ($T{=}1$), 
    we obtain partial sequences with the advantage of pruning a lot of data, allowing a faster learning speed. This pruning does not lower the performance of the Markov model since $T{=}1$. With $T{=}0$, the training set of the Markov model is composed of full sequences, there is no pruning. 
    \item Go to Step 3 with the newly selected subtask/action belonging to the task model and repeat until the end-user closes the interactive session.
\end{enumerate}

When a new interactive session is initiated, the LRS currently determined will go to Step 3  and repeat the process until the new session ends. \autoref{fig:adaptation-ex1} illustrates this process with a fictional task model decomposed into seven subtasks. During the first session, the first user enters the FUI that has been fractionally reified from $T1 ||| T2$ (parallel), executes the subtask contained in $T1$, then proceeds to $T3$ and $T4$ until returning to $T2$ in the first container.

Then, the end-user successively moves to $T6$, $T5$, $T7$, and back to $T5$ before closing the session. The resulting LRS is $T1$, $T3$, $T4$, $T2$, $T6$, $T5$, $T7$, $T5$, which is centralized to adapt the GUI for any future session. When the GUI is restarted (bottom part of \autoref{fig:adaptation-ex1}), $T1$ and $T3$ are subject to fractional reification since they begin the LRS. Similarly, $T4$ and $T2$ are offered. The user produces a new interaction sequence that automatically updates the LRS. We distinguish two scenarios depending on the continuity of sessions:
\begin{figure*}
    \centering
    \includegraphics[width=\textwidth]{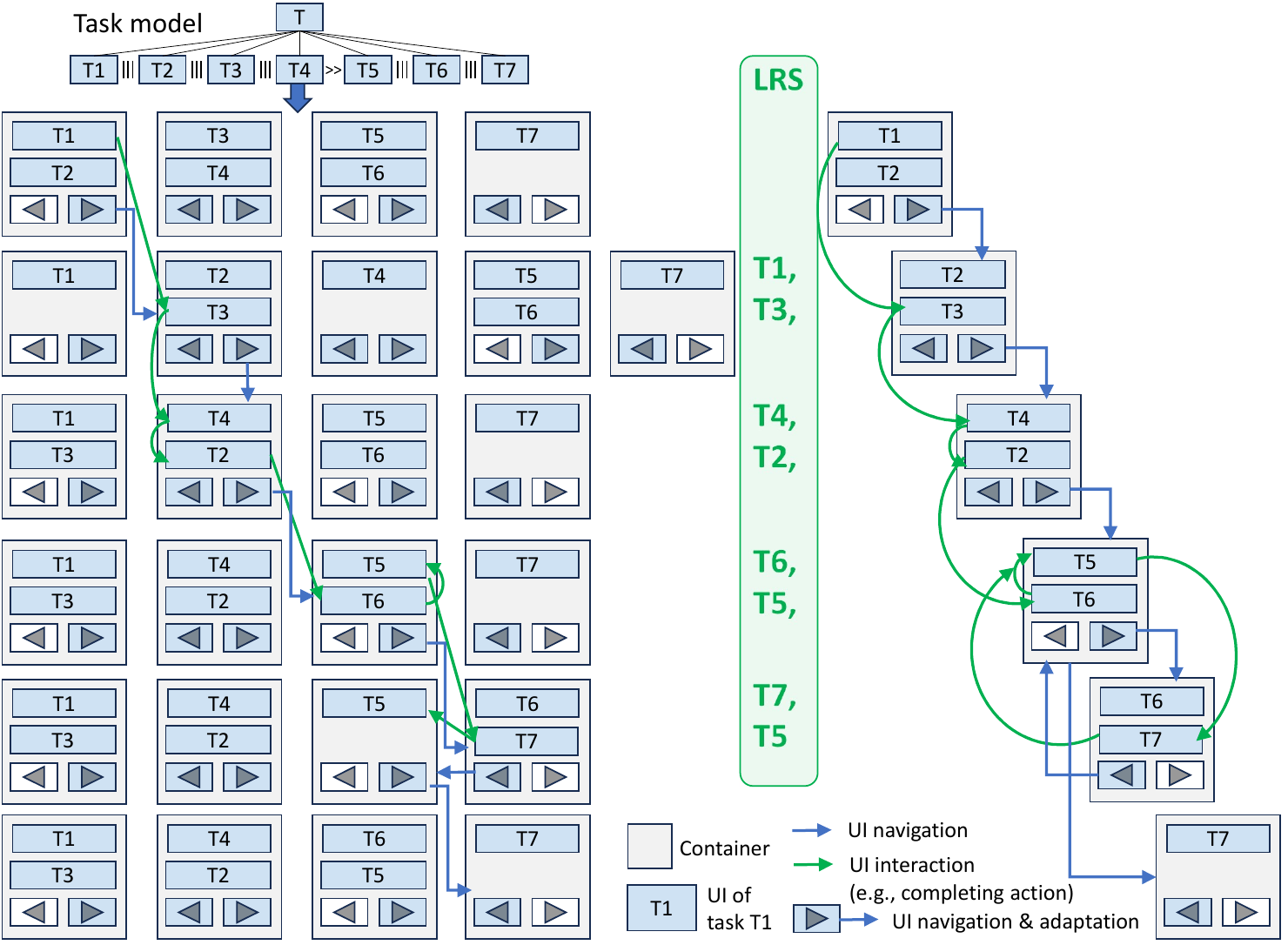}
    \vspace{-16pt}
    \caption{An example of GUI adaptation for both scenarios: intra-session (left) \vs inter-session scenario (right).}
    \label{fig:adaptation-ex2}
\end{figure*}

\begin{figure*}
  \centering
  \includegraphics[width=.97\linewidth]{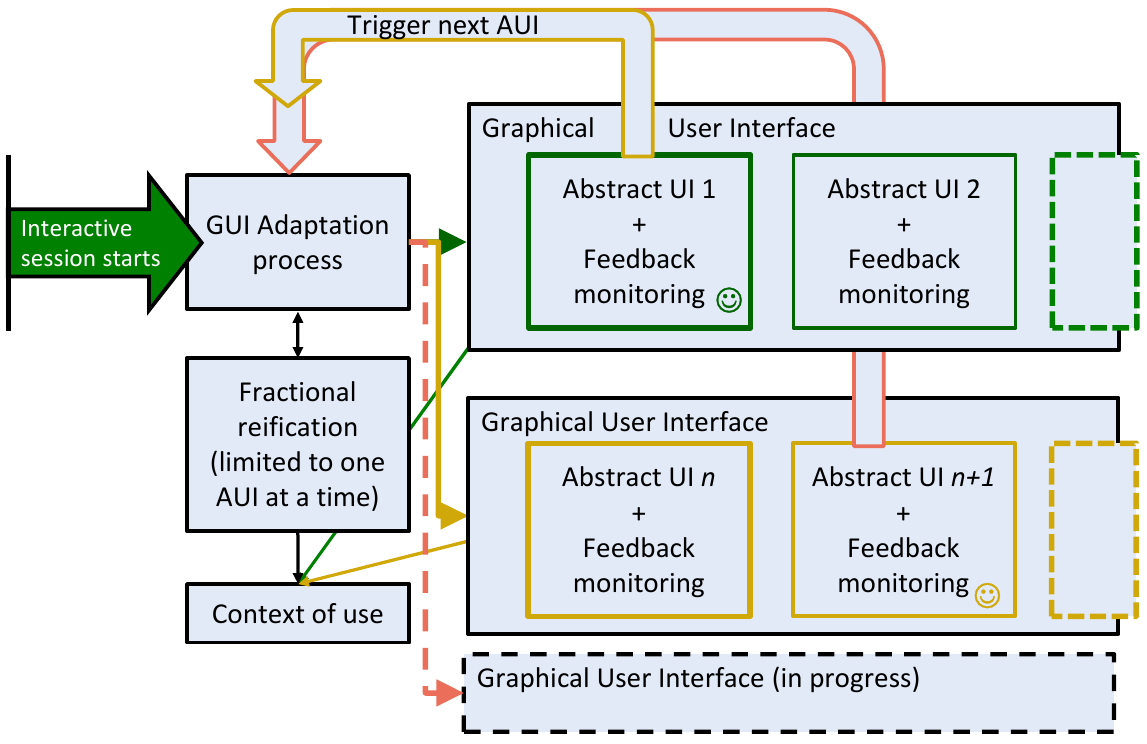}
  \vspace{-8pt}
  \caption{Workflow of the GUI adaptation in the inter-session scenario.}
  \label{fig:intersession-flow}
\end{figure*}

\begin{figure*}
  \centering
  \includegraphics[width=.97\linewidth]{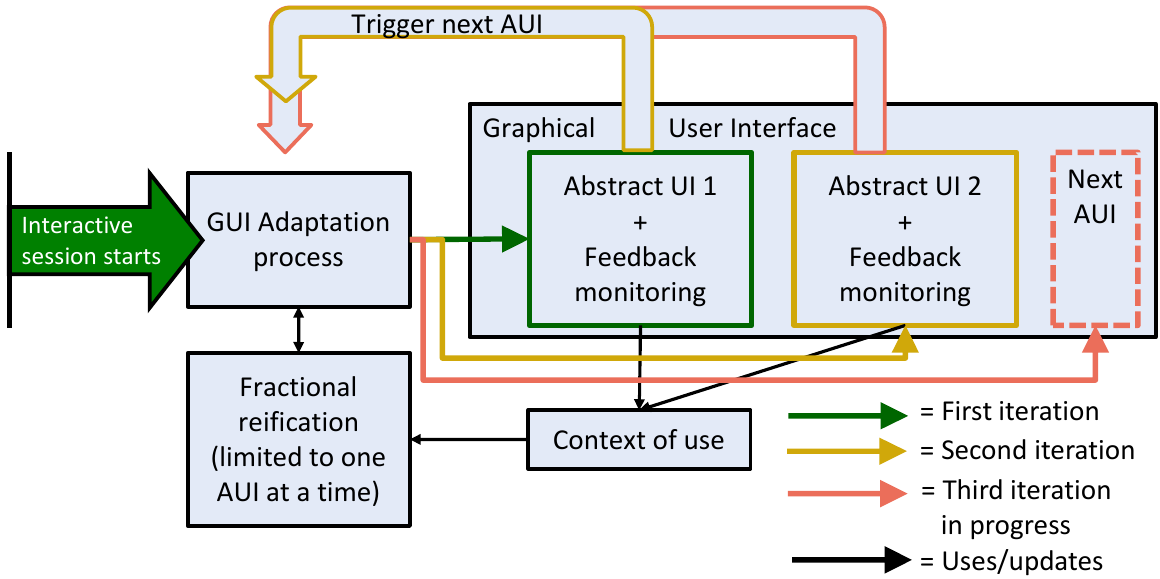}
    \vspace{-8pt}
  \caption{Workflow of the GUI adaptation in the intra-session scenario}
  \label{fig:intrasession-flow}
\end{figure*}

\begin{itemize}
    \item An \textit{inter-session scenario}: when the adaptation is automatically initiated by \tp each time a new interactive session is launched. The example above corresponds to this scenario with two sessions in \autoref{fig:adaptation-ex1} and in \autoref{fig:adaptation-ex2}, right. This workflow is shown in \autoref{fig:intersession-flow}.
    \item An \textit{intra-session scenario}: when the adaptation is manually initiated by the user and supported by \tp. \autoref{fig:adaptation-ex2} compares the GUIs adapted on the same LRS: the left part shows an adaptation of all AUIs performed at run-time during the same session, while the right part shows an adaptation of the currently being executed AUI only. This workflow is shown in \autoref{fig:intrasession-flow}. A single AUI is generated at a time: during the first session, AUI1 is generated (in green) with which the user is interacting, thus updating the LRS to trigger an adaptation the next time; during the second session, the adaptation generates AUI2 (in yellow), which will update the LRS for a new adaptation next session (in red). 
\end{itemize}

\begin{figure*}
    \centering
    \includegraphics[width=.8\textwidth]{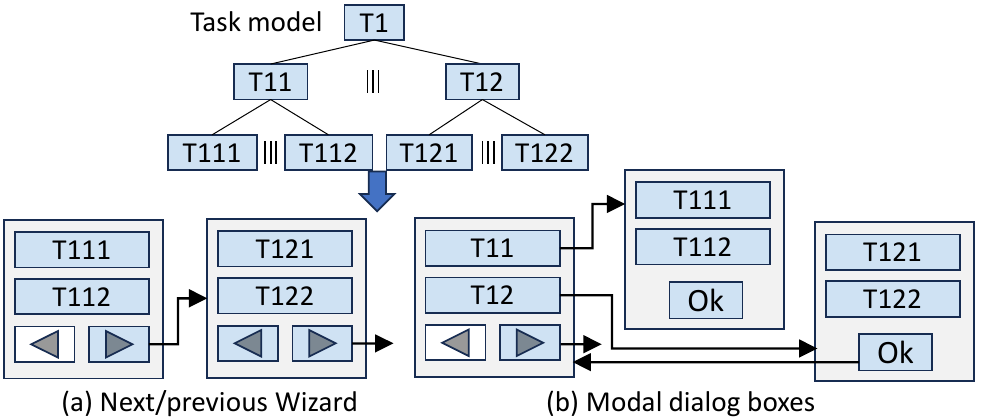}
    \caption{A first example of the GUI adaptation approach.}
    \label{fig:adaptation-ex3}
\end{figure*}

\begin{figure*}
\centering
	\includegraphics[width=.67\textwidth]{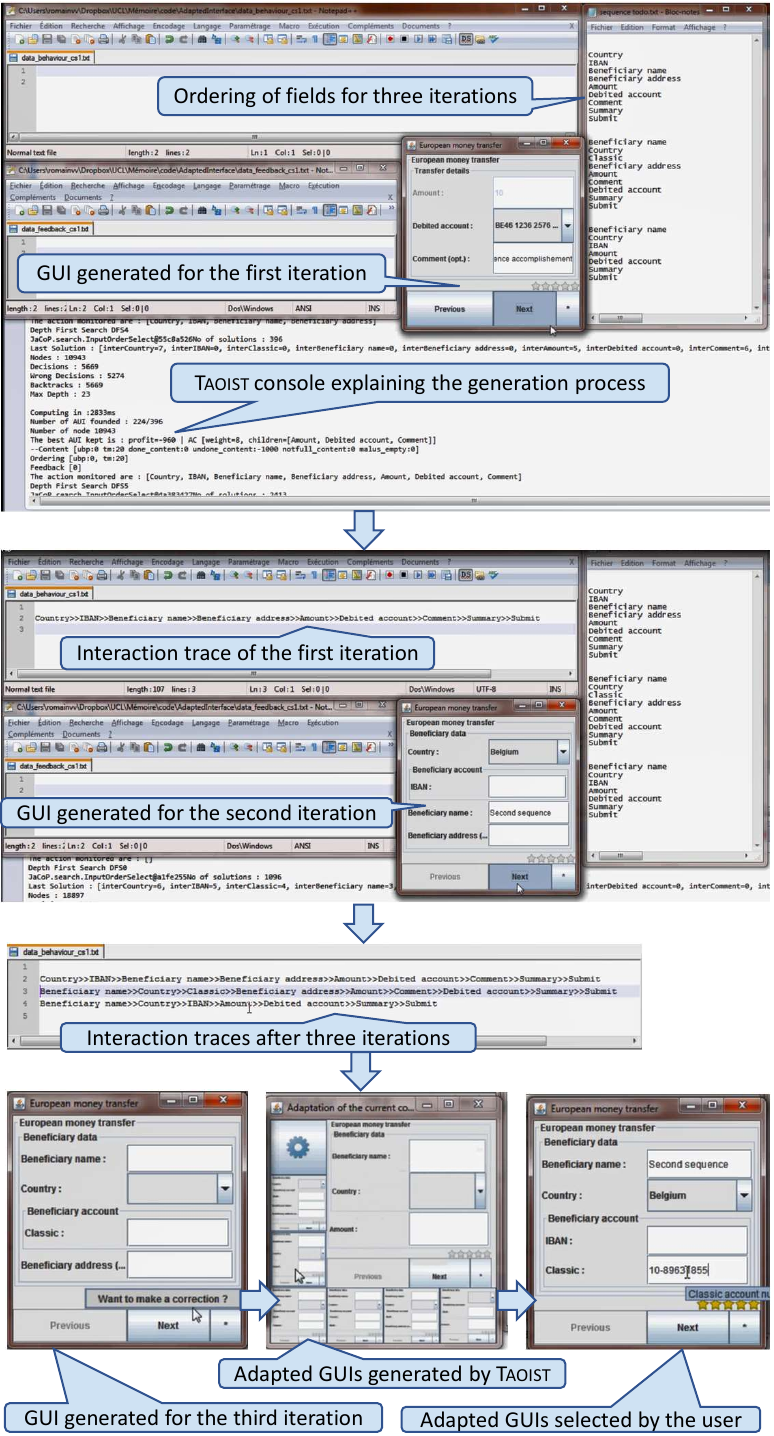}
	\caption{\tp walkthrough after three iterations.}
	\label{fig:taoist_walkthrough}
\end{figure*}

\autoref{fig:adaptation-ex3} shows two potential AUIs derived from the same task model (left): a typical wizard GUI with classical \texttt{Next} and \texttt{Previous} push buttons that are activated according to the constraints imposed by the temporal operators (center) and a GUI with behaviors opening a model dialog box each time a task is activated (right). This type of AUI generation mainly occurs in the intra-session scenario since all AUIs involved are available at a time, as represented in the left part of \autoref{fig:adaptation-ex2}. \autoref{fig:intrasession-flow} shows three adaptation iterations within the same session: a first iteration in green adapts the current AUI1 to the context of use by fractional reification and triggers a second iteration in yellow, which, in turn, triggers a third iteration in red, and so forth (\autoref{fig:taoist_walkthrough} for a walkthrough). \autoref{fig:intersession-flow} shows three adaptation iterations across two interactive sessions: a first iteration in green adapts the current abstract UI, similarly to the intra-session scenario, monitors and collects feedback from the end user; this feedback will be exploited for the next iteration in yellow, which takes place in another interactive session, and so forth. 


\subsection{Task Model}
To simplify steps (3) and (4) in adaptation, we adhere to the W3C recommendation to define a task model~\cite{W3C-Task:2014} (where four task categories are defined: abstract, interactive, manual and system) and the W3C recommendation to define an abstract UI~\cite{W3C-Abstract:2014} (where four types of abstract interaction components are considered: selection, input, output, and trigger -- see \autoref{fig:task-aui} in \autoref{app:metamodels} for more details). The rules listed in \autoref{tbl:task2aui} select an appropriate abstract interaction component depending on the attribute and transform them from the task model to an abstract UI. Navigation between these abstract interaction components is obtained using the Activity Chaining Graph~\cite{Luyten:2003}.

\begin{table*}[]
\centering
\begin{tabular}{|l|r|l|}
\hline
Attribute data type	& Property & Abstract Int. Component (AIC) \\
\hline
Boolean		& & Check button \\
Hour		& & Profiled edit field with regular expression \\
Date		& & Profiled edit field with regular expression\\
Char		& & Alphanumeric edit field\\
URL		& & Link\\
Hashtag		& & Profiled edit field with regular expression\\
Media		& & Browse push button linked to media manager\\
String	&30	&Single-line edit field\\
	&60	&Two-line edit field\\
		& & Multi-line edit field\\
Integer	& 2	& Slider\\
		& & Profiled edit field\\
Real	& &	Profiled edit field\\
Enumeration	& 3	& Group of radio buttons\\
	& 7	& List box\\
	& 30& Combo box\\
	& >30 &	Accumulator \\
\hline
Method	& direct	& Push button \\
	& indirect	& Edit field associated to a semantic function\\
\hline
\end{tabular}
\vspace{+4pt}
\caption{Rules to define an abstract interaction component based on the attribute.}
\label{tbl:task2aui}
\end{table*}

\begin{figure}
    \centering
    \includegraphics[width=\linewidth]{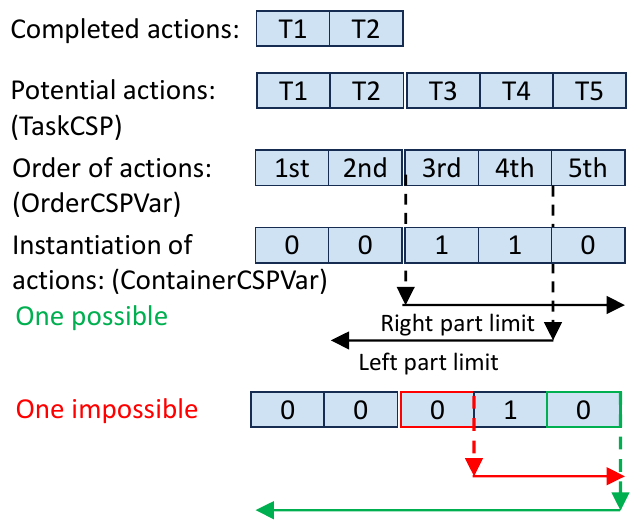}
    \vspace{-12pt}
    \caption{\tp: Fractional reification.}
    \label{fig:taoist_reification_scenario}
\end{figure}

To support a fractional reification (\autoref{fig:taoist_reification_scenario}),  the \texttt{ContainerCSPVar} variable specifies whether an action will appear in a container. The \texttt{GetPartialModels} method starts by initializing all variables of each action \texttt{TaskCSP}. In particular, \texttt{OrderCSPVar} and \texttt{Container CSPVar} are initialized so that the actions already executed by the user and the optional actions already displayed will remain in the left part of the LRS while keeping the other actions in the right part (\autoref{fig:taoist_reification_scenario}). Both parts are bounded by the first and the last actions of the container appearing in the LRS. The last instance in \autoref{fig:taoist_reification_scenario} violates this constraint because action \texttt{A3} should be in the right part. Once the fractional reification is performed, a FUI is generated to obtain a running GUI: the method \texttt{getPartialModel} creates a \texttt{JPanel Con-}  \texttt{tainer} that contains the widgets (\eg a check box, a radio button) automatically selected for each component (\autoref{tbl:task2aui}) with an identification label. The method then initializes a specific listener for the action monitoring. Next, the method \texttt{triggerEvaluation} adds the feedback features. The method \texttt{reifyPartialModelDialog} adds the navigation buttons between containers, such as \texttt{Go to Previous} and \texttt{Go to Next}. The fractional reification reduces the computational cost paid for real-time adaptation by generating only what is needed at a certain time for the end-user, not all possibilities.

\subsection{Markov Chains}

To create a $k^{th}$-order Markov model for predicting the next user's action  based on the $k$ previous and current actions, \tp creates a vocabulary $V$ containing actions of the task and determines action sequences from two sources:
\begin{itemize}
\item By converting the underlying tree of the task model into simulated sequences of actions based on the temporal operators (\eg sequence, concurrent, choice \cite{W3C-Task:2014}) to obtain the initial dataset.
 \item Monitoring the user's interaction, recording the sequence of actions monitored, and updating the LRS.
\end{itemize}

The $k^{th}$-order Markov model, once created, predicts the most probable action $t$ and compares it with the actual interaction history $h$.  For this purpose, we define an \textit{action score} as the probability of performing an action given that previous actions have been performed according to
\begin{equation}
    \label{eq:f1}
    \textrm{OrderFreeProbability(UI)} = \underaccent{S}{\max} \prod_{(t,h)\in S}P(t,h)
\end{equation}
where $S$ belongs to the set of simulated sequences of actions, and a \textit{content score} according to

ContentScorePrediction(UI) =
\begin{align}
    \label{eq:f2}
    \sum_{t}^{\text{tasks}}\left( \text{OrderIndepProbability}(t) \cdot \text{UBPWeight} \right)
\end{align}
where $t$ denotes any action or sub-task of a common parent task in the task model, $h$ denotes the interaction history of actions performed, and \texttt{UBPWeight} represents a weight of the probability. The \textit{task score} is defined as:

TaskScore(UserTask)=
\begin{align}
    \label{eq:f3}
    i \cdot \text{modelWeight} | \max_i \text{userTask} \left[0,i\right] \text{.sublist0}(\text{DFS}(tm))
\end{align}
where $\text{DFS}(tm)$ performs a depth-first search (DFS)~\cite{Tarjan:1971} in the task tree to generate a sequence of actions. The goal is to prioritize the action's location within the container, restricting the scope of predicted actions, therefore determining the position and order of the actions in the AUI. As a result, the score is more restrictive and increases with the degree of adequacy between the container's actions and the task model order. 
Variations of the \textsf{Ordering Score} can be calculated in addition to the \textsf{Content score}, such as evaluating the appropriateness of the AUI according to the task model in terms of order and hierarchy.

\section{Adaptation Examples}
\label{sec:case-studies}
This section describes three examples that illustrate how \tp supports GUI adaptation as described in \autoref{sec:taoist}. The three examples start from a task model in increasing order of their complexity to facilitate illustration and understanding.

\subsection{Example \#1}
\begin{figure*}
	\centering
	\includegraphics[width=.8\textwidth]{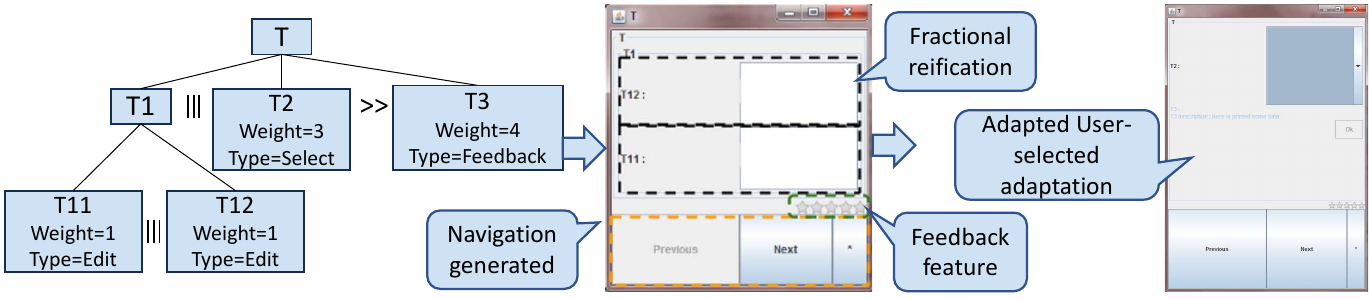}
	\caption{Adaptation of the example \#1.}
	\label{fig:example1}
\end{figure*}

Let us consider the task model illustrated in \autoref{fig:example1}, where the task $T$ is decomposed into three sub-tasks, $T_1$ (which is further decomposed into a parallelism between $T_{11}$ and $T_{12}$), $T_2$, and $T_3$. First, the reification process computes a score for all possible and displays the best-scored one. Let us assume that the user fills out the two text fields $T_{11}$ and $T_{12}$, resizes the window, and then clicks on the ``*'' feedback button. The reification re-evaluates the score of all partial AUIs, the ones composed of actions already displayed and filled have a score of malus. The prediction tool uses the performed actions ($T_{11}$, $T_{12}$) to query the Markov Chain and to generate possible adapted GUIs. The feedback feature enables the user to select the AUI containing the two last actions to be achieved: the widget $T_3$ is disabled in the selected adapted UI. When the user performs $T_2$ and starts the right part of the enabling operator by doing $T_3$, the other widgets are deactivated since already completed (\autoref{fig:example1}).

\begin{figure*}
	\centering
	\includegraphics[width=\linewidth]{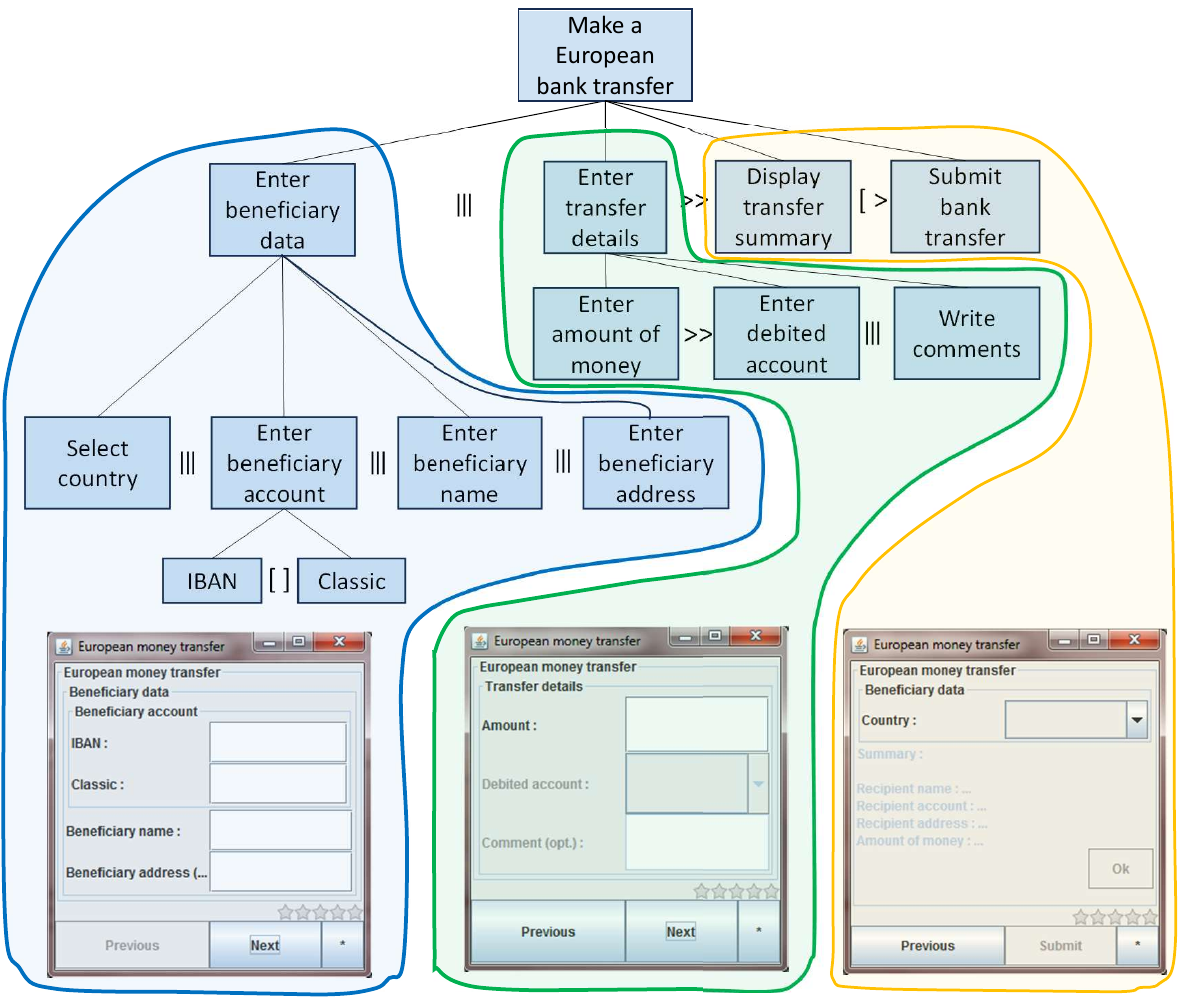}
        \vspace{-6pt}
	\caption{Task model and first GUI of the example \#2.}
	\label{fig:example2}
\end{figure*}

\subsection{Example \#2}
The second example concerns a bank transfer application for which there are many variations of the same task, whose model comprises three sub-tasks (\autoref{fig:example2}: the user must provide the beneficiary of the transfer, then specify the amount of money for this transfer, then select which bank account will be debited, which is determined after specifying the amount since the application checks if the account balance can accommodate the transfer. The user can optionally give some comments. A summary of the bank transfer is displayed before submission. The bottom part of \autoref{fig:example2} shows the different containers shown to the user the first time the interface is used, \ie without monitored data or feedback. We hypothesize that the user performs the following sequence of actions: \texttt{Country}, \texttt{IBAN}, \texttt{Beneficiary name},   \texttt{Beneficiary address}, \texttt{Amount}   

\texttt{Debited}, \texttt{account}, \texttt{Comment}, \texttt{Summary}, and \texttt{Submit}. 

The next time the user performs the same task, the processing of the previous sequence of actions will result in the first adaptation shown in \autoref{fig:example2-adapted}. According to this sequence, the costs of the first and second layouts are 46, 27, respectively. Let us now consider that the user executes a bank transfer of both types, \ie with a classic account number and with an \href{https://www.ibancalculator.com/}{International Bank Account Number (IBAN)}, but with different sequences each. The user fills out the optional fields for a classic transfer but does not fill them in the case of an IBAN transfer. The monitored sequences are thus the following ones, all being completed via \texttt{Submit}: 

\begin{figure*}
	\centering
	\includegraphics[width=.83\linewidth]{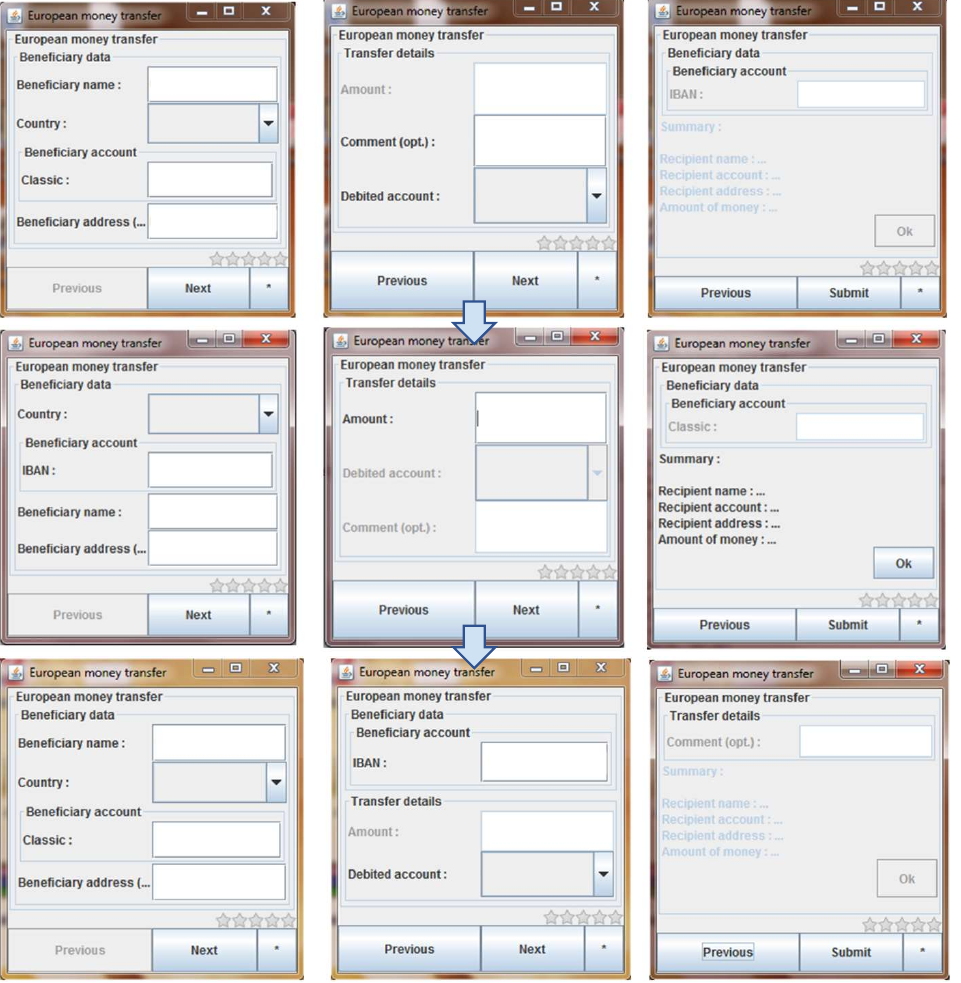}
	\caption{Adaptation of the example \#2.}
	\label{fig:example2-adapted}
\end{figure*}

\begin{itemize}
    \item \texttt{Country}, \texttt{IBAN}, \texttt{Beneficiary name}, \texttt{Beneficiary address}, \texttt{Amount}, \texttt{Debited account}, \texttt{Comment},   \\
     \texttt{Summary}  
    \item \texttt{Beneficiary name}, \texttt{Country}, \texttt{Classic}, \texttt{Beneficiary address}, \texttt{Amount}, \texttt{Comment}, \texttt{Debited account}, \\ \texttt{Summary}
    \item \texttt{Beneficiary name}, \texttt{Country}, \texttt{IBAN}, \texttt{Amount}, \texttt{Debited account}, \texttt{Summary}
    \item \texttt{Beneficiary name}, \texttt{Country}, \texttt{Classic}, \texttt{Beneficiary address}, \texttt{Amount}, \texttt{Comment}, \texttt{Debited account}, \texttt{Summary}
    \item \texttt{Beneficiary name}, \texttt{Country}, \texttt{IBAN}, \texttt{Amount}, \texttt{Debited account}, \texttt{Summary}
\end{itemize}

Exploiting these new sequences of actions, \tp generates the GUIs reproduced in the second and third row of \autoref{fig:example2-adapted} for classic and IBAN transfers, respectively. Depending on the type of account selected, the next container displays \texttt{Amount} and then either \texttt{Comment} or \texttt{Debited account}, which was made possible by the use of second-order Markov chain. A first-order Markov chain is unable to make this distinction because the history would be composed of the single action \texttt{Amount}. Instead, the sequence “\texttt{Classic}, \texttt{Beneficiary address}, \texttt{Amount} or \texttt{IBAN}, \texttt{Amount}, thus increasing the probability of displaying \texttt{Comment} or \texttt{Debited account}.
\begin{figure*}
	\centering
	\includegraphics[width=.97\linewidth]{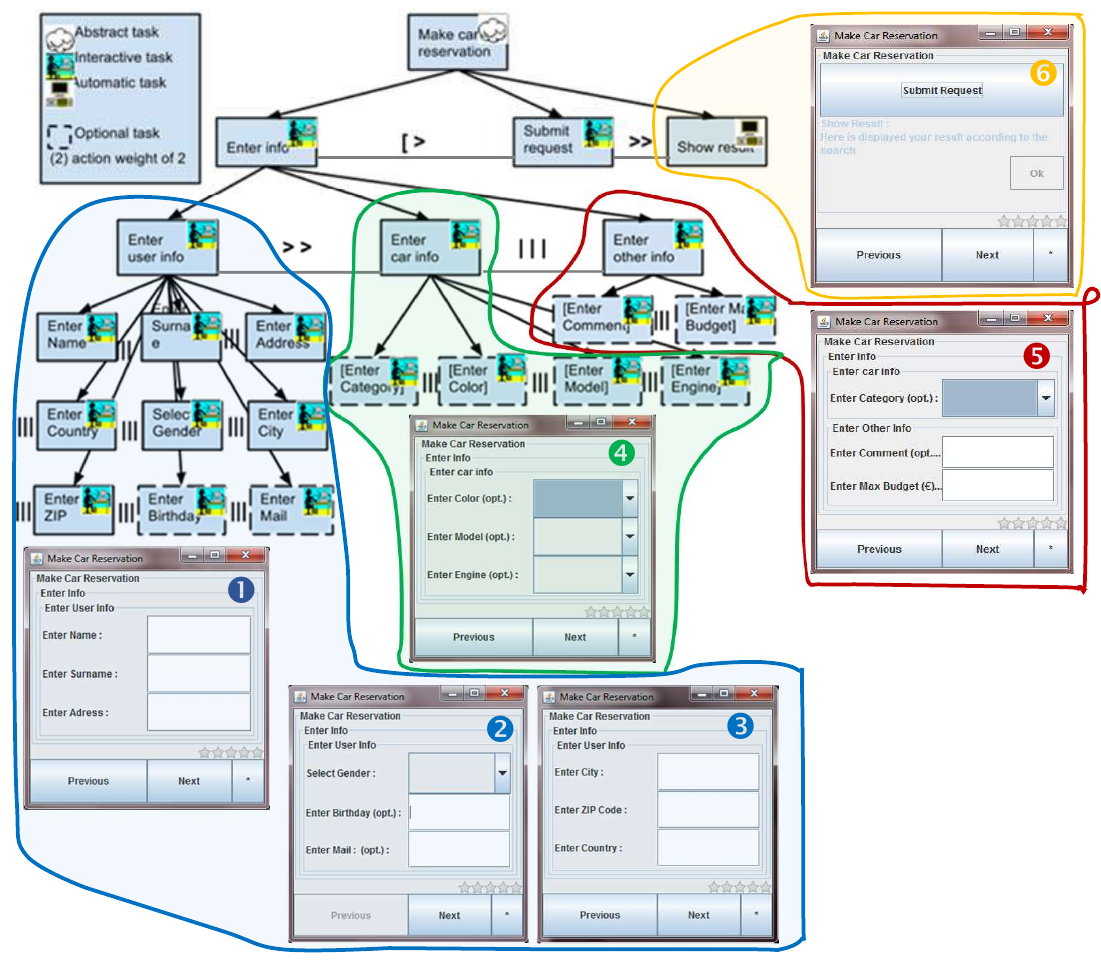}
	\vspace{-8pt}
	\caption{Task model of the ``Car rental'' W3C reference case study~\cite{W3C-Task:2014} and its corresponding initial GUIs.}
	\label{fig:task2AUI}
\end{figure*}
Optional actions are always displayed once and accessible through the \texttt{Next} push button, even if they are unlikely to happen.  Optional actions are usually presented before an enabling or disabling operator so that the user does not forget them. In this case, the layout appropriateness~\cite{Sears:1993} would be better without displaying the optional actions in the containers, but this would produce an incomplete UI. We also observe that the \texttt{IBAN} and \texttt{Classic} fields are not displayed together on the first container, although the above sequences suggest the opposite. This is an adverse effect of the score normalization according to the number of actions in the container. However, this can always be solved through user feedback.

\subsection{Example \#3: The W3C Reference Case Study}

\begin{figure*}
    \centering
    \includegraphics[width=\textwidth]{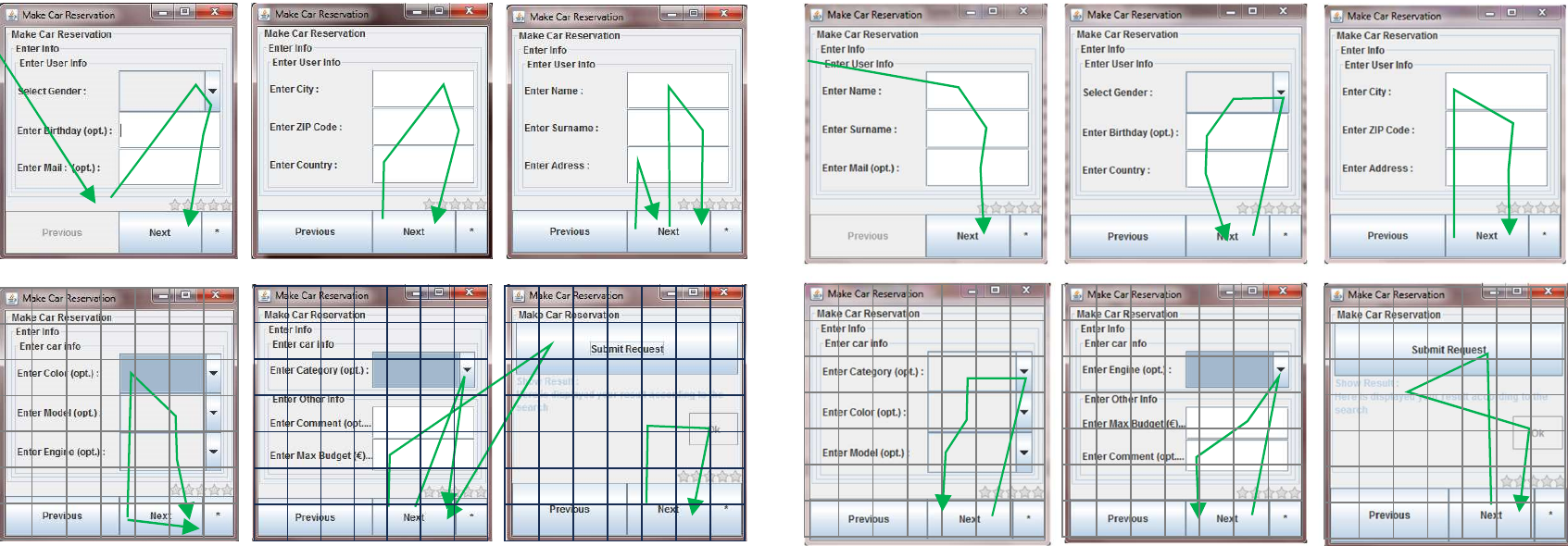}
    \caption{Layout appropriateness: (left) initial GUI: $L{=}67$, (right) adapted GUI: $L{=}49$.}
    \label{fig:layout}
\end{figure*}
To further exemplify the adaptation by \tp, we selected the W3C reference case study for adaptation~\cite{W3C-Task:2014}: its task model describes a car rental application allowing users to select various car characteristics, to specify user information (\eg gender, birthday, email, address), to enter the budget, and to leave comments, before launching the search. Applying the adaptation outlined in \autoref{sec:process} generates 5 AUIs corresponding to the branches of the task model (\autoref{fig:task2AUI}). Let us consider the following sequence of user's actions performed on the initial GUI reproduced in \autoref{fig:task2AUI}: \texttt{Enter Name}, \texttt{Enter Surname}, \texttt{Enter Mail}, \texttt{Select}  \texttt{Gender}, \texttt{Enter Birthday}, \texttt{Enter Country}, \texttt{Enter City}, \texttt{Enter zip Code}, \texttt{Enter Address}, \texttt{Enter Category}, \texttt{Enter Color}, \texttt{Enter Model}, \texttt{Enter Engine}, \texttt{Enter Max Budget}, \texttt{Leave Comment}, \texttt{Submit Request}, and \texttt{Show Result}. Each time a new sequence of actions is recorded, we compute the Layout Appropriateness~\cite{Sears:1993}, a metric expressing the complexity of a layout depending on the distance traveled between widgets used by the actions in it.  Based on the first sequence, the layout appropriateness is $L{=}67$ (\autoref{fig:layout}-left).

\tp then adapts the initial AUI to this first sequence to obtain a new set of AUIs. Another sequence of actions on this adapted GUI gives a layout appropriateness of $L{=}49$ (\autoref{fig:layout}-right), which proves that the adaptation reduces this metric. Now we suppose that another user only fills in the mandatory fields and the \texttt{Enter Max Budget} field to record the following sequences:
(1) \texttt{Enter Name}, \texttt{Enter Surname}, \texttt{Enter Mail}, \texttt{Select Gender}, \texttt{Enter Birthday}, \texttt{Enter Country}, \texttt{Enter City}, \texttt{Enter zip code}, \texttt{Enter Address}, \texttt{Enter Category}, \texttt{Enter Color}, \texttt{Enter Model}, \texttt{Enter Engine}, \texttt{Enter Max Budget}, \texttt{Enter Comment}, \texttt{Submit Request}, \texttt{Show Result};
(2)	\texttt{Enter Name}, \texttt{Enter Surname}, \texttt{Select Gender}, \texttt{Enter Country}, \texttt{Enter City}, \texttt{Enter zip Code}, \texttt{Enter Address}, \texttt{Enter Max Budget}, \texttt{Submit Request}, \texttt{Show Result};
(3)	\texttt{Enter Name}, \texttt{Enter Surname}, \texttt{Select Gender}, \texttt{Enter Country}, \texttt{Enter City}, \texttt{Enter zip Code}, \texttt{Enter Address}, \texttt{Enter Max Budget}, \texttt{Submit Request}, \texttt{Show Result}.
\begin{figure*}
    \centering
    \includegraphics[width=\textwidth]{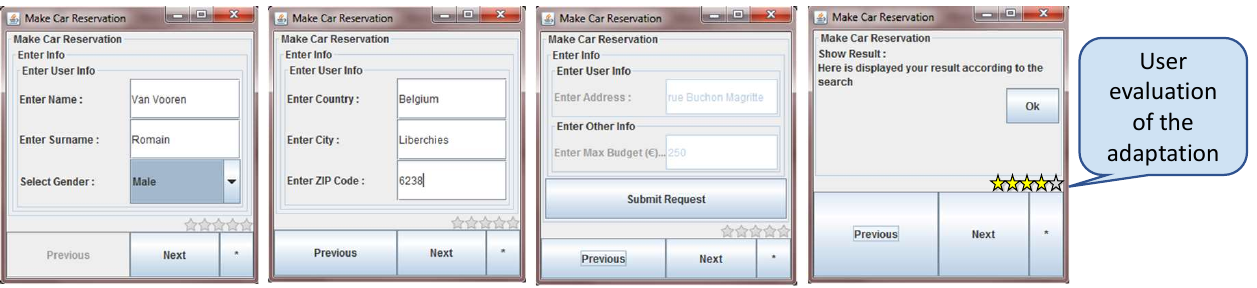}
    \caption{Final AUI after multiple adaptations.}
    \label{fig:finalAUI}
\end{figure*}

\begin{figure}
	\centering
	\includegraphics[width=\linewidth]{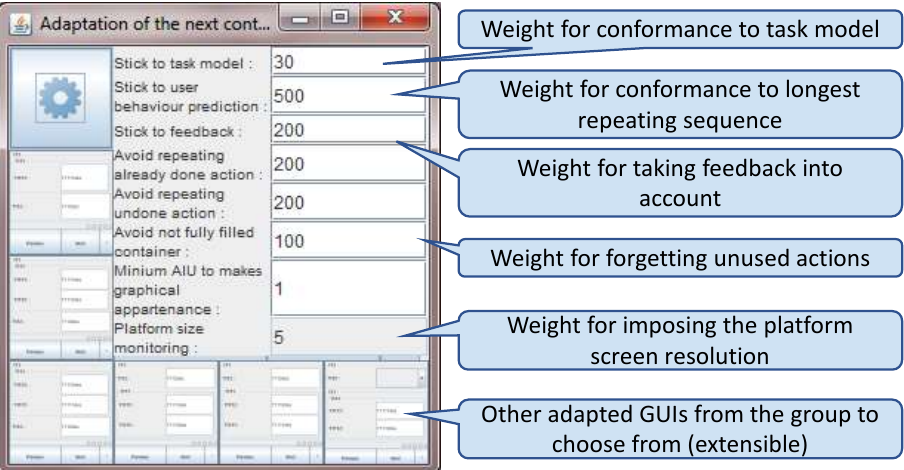}
	\caption{Controllability of machine learning features.}
	\label{fig:controllability}
\end{figure}

Despite the shortcuts made by this last user in the two last sequences by avoiding filling optional fields, the \texttt{Submit} push button would not show up earlier: we always print optional actions with a low probability at least once so that the user would not forget that they exist. Not showing optional actions could be intuitive when the left part of the disabling operator is disabled by clicking on a button. If we allow printing optional actions less than once and if a low weight is assigned to the score malus given to the AUI displaying already shown actions, the button \texttt{Submit Request} would appear several times in the containers and will show up earlier. According to the last observed sequences, this push button would be on the third and the last AUI (\autoref{fig:finalAUI}). The user is not forced to skip the containers showing the optional tasks dealing with the \texttt{Car info} sub-task: the \texttt{Submit Request} button can be pushed, then \texttt{Next}, and the following AUI will contain the query result. If the optional actions should be performed, there is no need to push the \texttt{Submit Request} button and the next AUI will display the optional actions. In any intra-session or inter-session scenario, the end user can provide \tp with some feedback on the last adaptation by rating it: the rightmost AUI of \autoref{fig:finalAUI} shows a rating of 4/5, which will be considered in the future. Furthermore, the end user can, at any time, control the adaptation by specifying several parameters to control the adaptation (\autoref{fig:controllability}), such as weighting to what extent the structure of the initial task model should be preserved while adapting the GUI or weighting to what extent the LRS obtained by previous interactions with the group should be used (in this case, only personal adaptations will be recorded to an identified user). The bottom part of \autoref{fig:controllability} includes a set of other adaptations obtained by other members of the group based on the LRS: the end user can then choose one of the adaptations.

\section{Performance Evaluation}
\label{sec:performance}
\begin{figure}
	\centering
	\includegraphics[width=\linewidth]{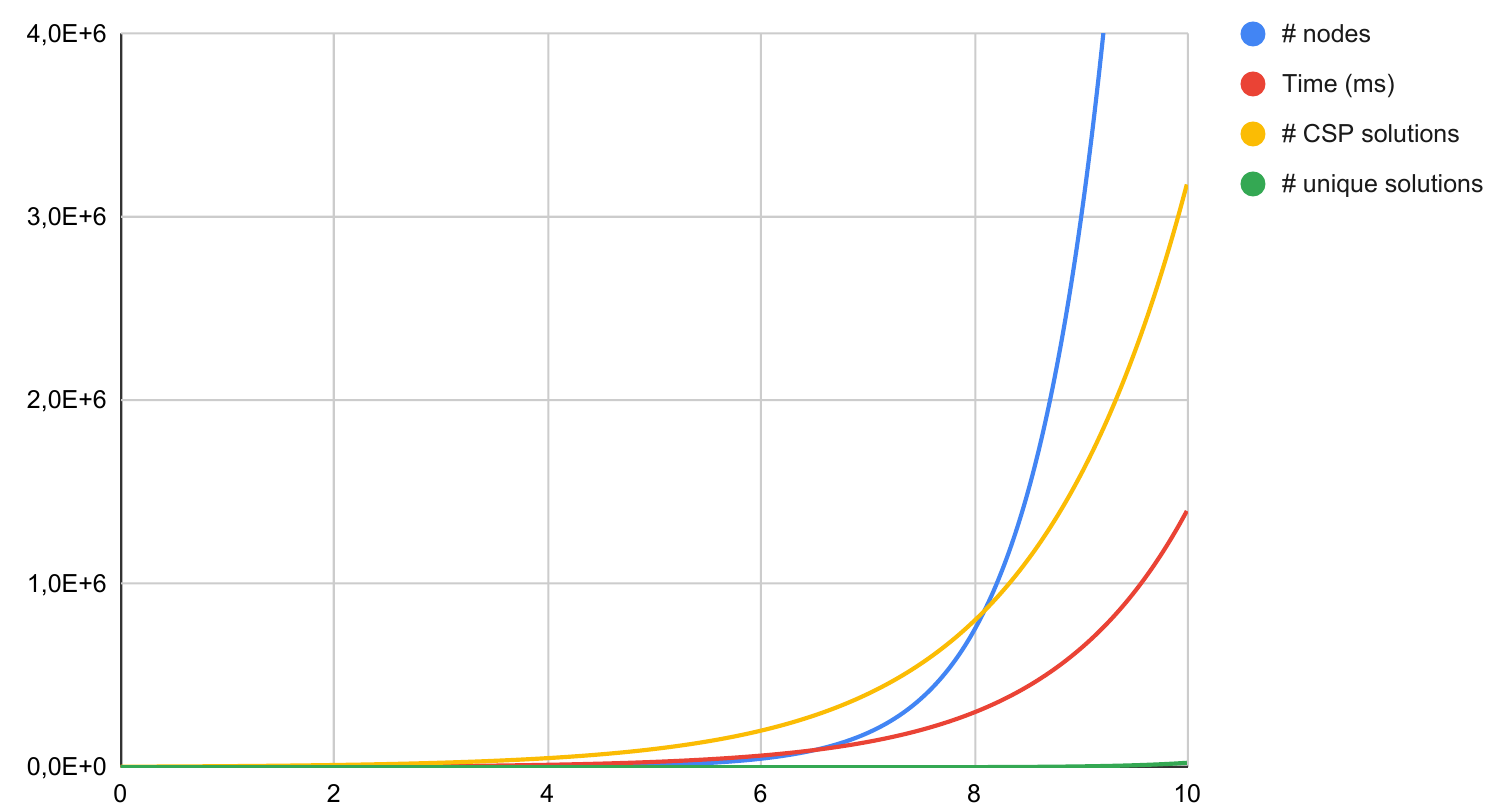}
	\caption{Number of nodes, execution time, and number of solutions depending on the concurrent actions.}
	\label{fig:taoist_caseStudy3_performanceGraph}
\end{figure}

The task model used in \autoref{fig:task2AUI} is flexible in offering concurrency operators (``|||'' in \autoref{fig:task2AUI}) between actions, thus enabling the user to perform the sub-tasks and actions in parallel or in any order, such as within parts \ding{202} to \ding{204}. This flexibility induces a combinatorial explosion in the number of ways and orders to enter the data and execute actions into sequences since no constraint is imposed for any sequence. \autoref{fig:taoist_caseStudy3_performanceGraph} illustrates this combinatorial explosion by showing the number of nodes, the execution time, and the number of possible action sequences depending on the number of concurrent tasks. The plot has been made with a platform weight of 4 and an action weight of 1.

To reduce the number of solutions and to ensure the tractability of the problem, we transform the concurrent operators into sequential operators between several tasks. This affects the task structure, which should be maintained to some extent depending on the weight for task conformance specified by the end user (\autoref{fig:controllability}).

\begin{figure*}
	\centering
	\includegraphics[width=\textwidth]{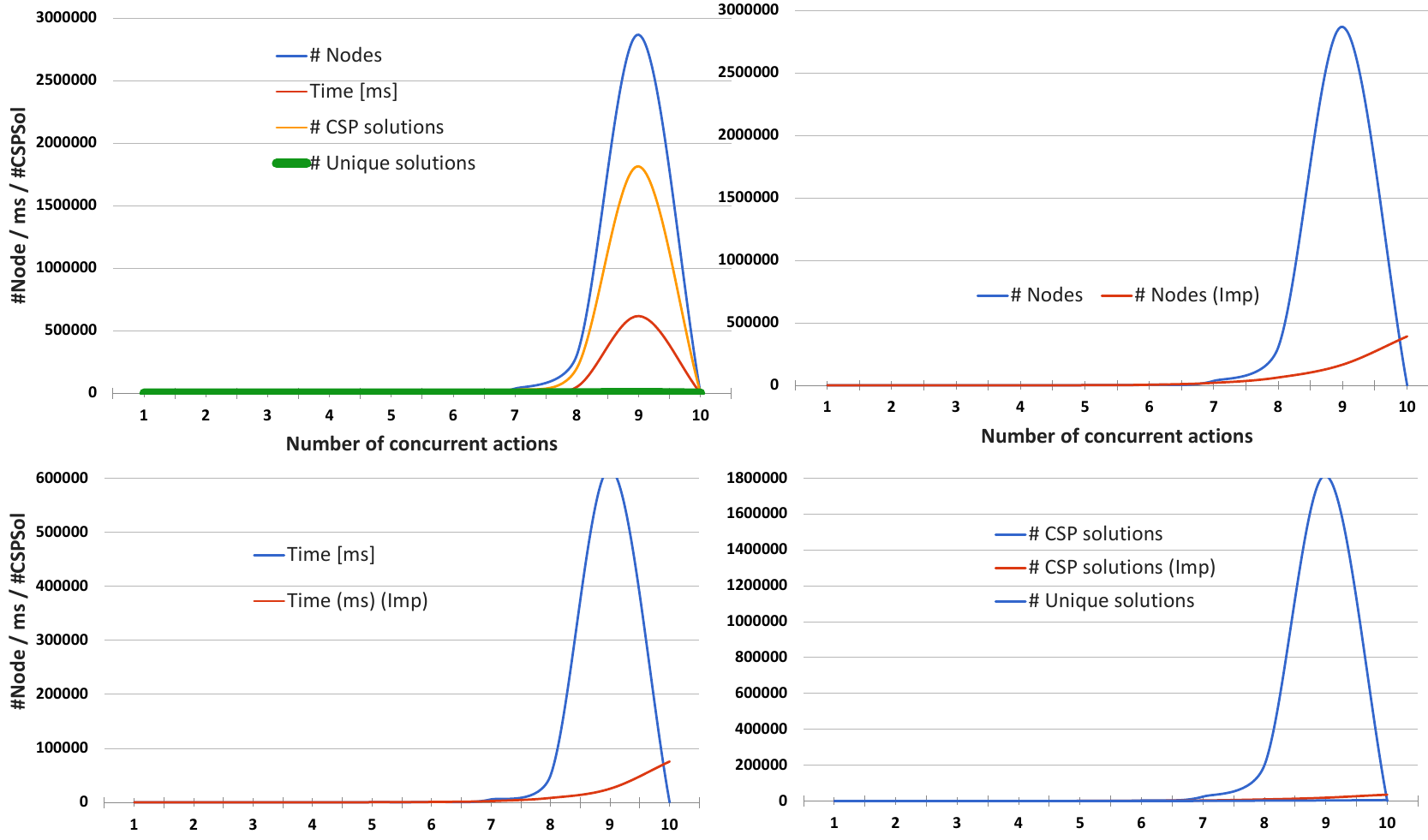}
	\caption{\tp Performance schemes: concurrent actions (top left), evolution of nodes with improvement (top right), evolution of time with improvement (bottom left), evolution of the number of solutions with improvement (bottom left).}
    \label{fig:performance}
\end{figure*}

 Since \tp may infer AUI solutions that are the same, which are referred to as \textit{duplicates}, we distinguish the number of unique solutions (\#Solution unique) and the number of CSP solutions (\#CSP solutions). Removing these duplicates already reduces the total number of CSP solutions, but not significantly. For this purpose, we introduced a \textit{partial search} within the task tree by Tabu Search, where we are not forced to expand all possible branches of the task tree when we do not want to find all the solutions. With this partial search, we need to use the score to guide the search to find only the $k$ best solutions. In this way, we ensure that the $k$ solutions found will be the best ones because the search is incomplete and non-optimal.  \autoref{fig:performance} shows the improvements in performance brought about by adding a new CSP variable for each action. While a significant performance gain is obtained, the complexity is still exponential. 

\section{Quantitative and Qualitative Evaluation}
\label{sec:qualitative}
\begin{figure*}
	\centering
	\includegraphics[width=\textwidth]{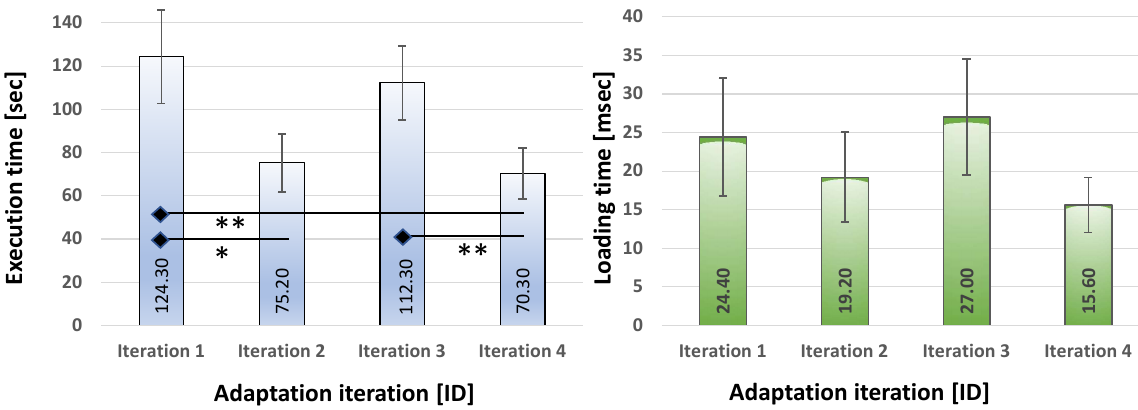}
	\vspace{-8pt}
	\caption{Execution time (left) and \tp loading time (right) over four adaptation iterations.}
    \label{fig:iterations}
	\vspace{+8pt}
\end{figure*}

To evaluate the adaptation supported by \tp, we also conducted a qualitative evaluation. Based on a random sampling, we recruited 10 participants (4 female, 6 male) from an international mailing list of UI/UX practitioners from our network of collaborators.
An email was sent to the mailing list asking practitioners willing to assess \tp. No compensation was offered since practitioners all volunteered.
After filling in a GDPR-compliant demographic questionnaire, participants were asked to carry out four times the same ``Car rental'' task (\autoref{fig:task2AUI}) with \tp running in an inter-session scenario, each time with a different set of input data. 
No particular order was specified, thus allowing the participants to perform sub-tasks and actions in any order allowed by the task model and reflected in \tp.  After each task is completed, \tp triggers an adaptation that is applied for the next task until the fourth task is completed. 
We measured the total time for each task completion (\autoref{fig:iterations}-left) and the \tp loading time required to perform the adaptation and to re-launch the GUI (\autoref{fig:iterations}-right) across sessions.

\subsection{Threats to Validity}
\begin{figure*}
	\centering
	\includegraphics[width=\textwidth]{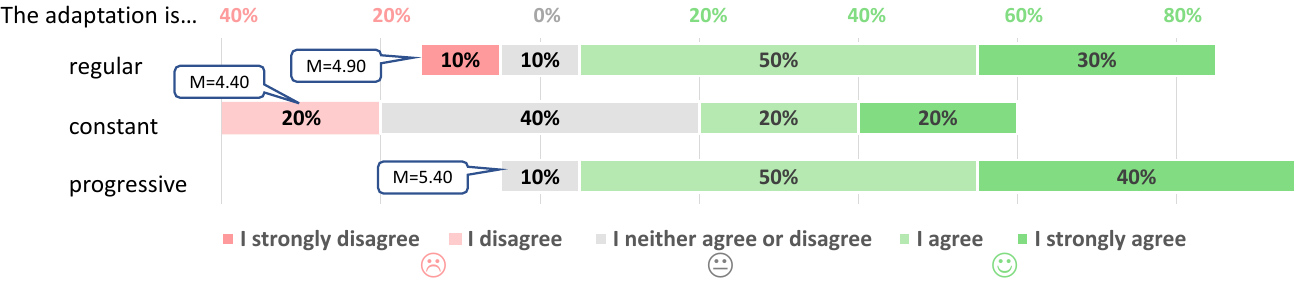}
	\vspace{-16pt}
	\caption{Distribution of participants' answers for the three adaptation properties.}
    \label{fig:evaluation}
\end{figure*}

 We ran a Kruskall-Wallis test with post-hoc Nemenyi pair-wise comparisons to compare the differences in execution times between the four iterations ($n{=}40$). The differences were statistically significant: $H(3)=16.19$, $p{=}.00103^{**}$. Iteration 2 is significantly shorter than iteration 1 ($R{=}15.05$, $SD{=}3.69$, $p{=}.0209^{*}$), iteration 4 is significantly shorter than iteration 3 ($p{=}.0029^{**}$), and than iteration 1 ($p{=}.0066^{**}$). These results suggest that the first adaptation reduced the completion time compared to the starting condition and that the last adaptation did the same compared to the previous one. However, the third iteration had a completion time slightly lower than the starting time, but not significantly. A potential explanation for this change is that participants were probably confused during the third iteration by another unexpected event or in a situation of disruption~\cite{Hui:2009}. After a cognitive destabilization, participants go back to a re-stabilization that leads them to a final shorter completion. There was no significant difference between iterations 2 and 4, but the final state is shorter than the initial state. A Kruskall-Wallis test on the startup time does not reveal any difference ($H(3){=}6.09$, $p{=}.106$, \textit{n.s.}), although the various times follow the same trend as for the completion time.

At the end of the experiment, we asked participants to rate on a 7-point Likert scale to what extent they perceived the adaptation to be regular, constant, and progressive, as devised in the introduction. \autoref{fig:evaluation} shows that 80\% of participants agreed to perceive the adaptation as regular ($M{=}4.90^{*}$, significantly higher than the median, with only one participant reporting not having perceived any change) and even more as progressive (90\%, $M{=}5.40^{**}$, significantly high than median), but not as constant as expected (60\% of participants were not convinced, $M{=}4.18$, \textit{n.s.}).

\section{Benefits, Limitations, and Threats to Validity}
\label{sec:adv-lim}

\subsection{Benefits and Limitations}
The experience gained with the Markov Chains and the Longest Repeating Sequences shows the feasibility of a progressive and controllable adaptation. To automate as much as possible the process, \tp starts from a task model expressed according to W3C recommendations for task modeling~\cite{W3C-Task:2014} and abstract user interface~\cite{W3C-Abstract:2014}. While these recommendations are supported in the adaptation, the process is entirely governed by its advantages (\eg the various temporal operators and the ability to generate many abstract UIs based on the models), but is also limited by its expressiveness. This capability makes the adaptation particularly suitable for interactive information systems~\cite{Pastor:2007}, that are characterized by their transactional input and output and their typical operations such as create, read, delete, update, and search methods. While the adaptation process remains in principle independent of any implementation, the way \tp supports it remains limited by its expressiveness. Expanding the models with other attributes and concepts, and the generation rules (such as those listed in \ref{tbl:task2aui}) is still feasible, but remains unsupported by \tp.

The structure of the task model, particularly when it includes many flexible operators like concurrency, independent order, or choice, generates a combinatorial explosion of possibilities. A partial search, oriented towards finding a limited number of solutions, provides a feasible solution but not an optimal one. More interestingly, at each adaptation, the end user can give feedback via a rating bar and/or set some parameters that are useful for the adaptation. Among these parameters, the user can impose or release the history of the adapted interfaces for other users, belonging to the same group or not: the result is a display of the interfaces previously adapted for other members of the group among which the end user can make a new choice. Then, at any time, the proposed adaptation can be accepted, rejected, or taken among the alternatives. In this way, it is expected that GUI adaptations will become regular, constant, and progressive.

Concerning feedback, all the data captured from interaction might adjust the container concerned by the feedback, or influence more globally the adaptation process. Moreover, we may consider the user behavior prediction to predict the content of the widget based on the previous data filled by the user. This kind of tool looks like an enhanced auto-compilation tool. It could be interesting to deal with some kind of data processing/filtering adaptation mixed with container content adaptation. Right now, we are generating all the possible solutions, and then we select the best according to the score. It would be efficient to improve an existing solution with a local search technique to reach a better one without generating all the solutions by brute force like we did. A local search algorithm can be deployed for that. The algorithm was validated via a case study and results were evaluated and discussed according to the Layout appropriateness metric and a user evaluation. Performance and usability results demonstrate that there are significant benefits in predicting future states of user interaction. The results of the usability survey \cite{Beirekdar:2002,Beirekdar:2005} show that users perceive a system more useful when it follows their preferences.

\textbf{Threats to External Validity}.
For the quantitative and qualitative evaluation, we adopted a random sampling which is aimed at producing diverse samples, so this part of our evaluation can
not claim generalizability. Instead, we aimed to provide insights
into how practitioners would assess the regular, constant, and progressive characters of GUI adaptation, as supported by \tp. Future
experiments should confirm these insights with more diverse populations of practitioners and end users in other contexts of use. For this purpose, we could conduct a similar experiment with a stratified sampling based on the nine development expertise levels \cite{Costabile:2008} as used in \textsc{MoCaDix} \cite{Vanderdonckt:2019}, or with a cluster sampling based on personal traits \cite{Gajos:2017}.
Secondly, our evaluation is based on a single information system (a car rental application chosen for its simplicity and based on a task model), therefore potentially leading to disparities between the adaptations obtained for this particular case study and other real-world information systems. For the same reasons, we cannot generalize to other interactive applications than information systems. Other families of applications, such as safety-critical applications, control applications, and office systems, are not so appropriate candidates for adaptations.
Thirdly, our experiment took place in our laboratory with the same equipment to ensure control and consistency of the experimental conditions. Nothing prevents us to reproduce the experiment on the very right context of use of the participants with the platform they are used to have. Thus, the ecological validity is medium.

\textbf{Threats to Internal Validity}.
Our participants may have found the three questions about regular, constant and progressive adaptation somewhat difficult to grasp and to differentiate. We tried to mitigate this aspect by explaining them their difference independently of any particular case, based on \autoref{fig:requirements}. To ensure the natural character of the experiment, we did not impose any time constraint on participants to preserve the task completion time. Although a significant difference was found over time, this difference could be also attributed to any carry-over effect, such as a learning effect. However, since \tp performed some different adaptation for each iteration, the learning effect could be different.

\textbf{Threats to construct validity}. Execution time and loading time were used as quantitative measures. A 3-item questionnaire was used as a qualitative measure. Task completion rate is not reported since all participants completed the tasks for each iteration. Task error rate and subjective satisfaction after each iteration could have been used also. We opted for a simple procedure not to overwhelm the participants. A between-subject design could also be used instead of a within-subject.

\section{Conclusion and Future Work}
\label{sec:conclusion}
We presented \tp, a novel model-based approach to GUI adaptation that combines a task model and Markov Chains with the longest repeating Subsequences mechanism that was created to assess whether adaptation can be improved if regular, constant, and progressive.
The state space of the task interaction is discretely produced by the task model and the interaction observations are dynamically generated from a categorical distribution exploiting the longest repeating subsequences in the task model. 
A GUI is therefore automatically generated at runtime, a decisive advantage for pre-generating the action sequences that will populate the Markov Chain to calculate the longest repeating sequence. 
A task model explicitly defines the temporal order in which actions, such as data entries, displays, and function triggers, can take place. 
Future work should consider other approaches to deal with sequentiality constraints, perhaps less explicitly.

The weight assigned to the conformance with the task model can be adjusted by the end user to the extent that constraints imposed by this model can be relaxed to decrease the number of possible generations of future adapted interfaces (\eg a concurrency can be degraded into a sequence, a choice can be transformed into a sequence).

\tp supports both intra-session and inter-session adaptation scenarios, to be decided by end users. However, the conditions under which a participant first selects one of these two scenarios are not yet known. A participant may very well start with an intra-session scenario if adapting the interface to some needs and preferences, or with an inter-session scenario if adaptations already made elsewhere, especially by other group members, can help. This technique is suitable for supporting the transition of a novice user, who wishes to build on the experience capitalized by other users, to a more experienced user. The same participant can switch from one scenario to another depending on how the adaptation quality is evaluated and interpreted~\cite{Dessart:2011}, the two last stages of an adaptation life cycle~\cite{Abrahao:2021,Lopez:2007}. The ability of the user to evaluate this quality and to modify the parameters accordingly is key.

\begin{acks}
This work is supported by the EU European Innovation Council (EIC) Pathfinder-Awareness Inside challenge "\href{http://symbiotik-infovis.eu/}{Symbiotik}" project (1 Oct. 2022-30 Sept. 2026) under Grant no. 101071147. We also thank ROmain Van Vooren for his contribution to this work.
\end{acks}

\bibliographystyle{plainnat}
\bibliography{bibliography}

\clearpage
\newpage
\appendix
\section*{Appendix A. Implementation of \tp}
\label{app:implementation}
This appendix specifies, describes, and explains how \tp has been implemented to combine the two approaches: the task model-based approach and the Longest Repeating Subsequences on Markov Chains.
The software architecture of \tp aims to be as general as possible  to instantiate it for different adaptation scenarios. We use abstract classes and interfaces as much as possible to be generic and let the possibility to extend the framework by improving the existing features or adding new features. 

The three top abstract classes of \tp correspond to three high level concepts:
\textsf{Model}, \textsf{Reification}, and \textsf{AdaptingInterface} (\autoref{fig:top}). The first two abstract classes \textsf{Model} and \textsf{Reification} are based on the Cameleon Reference Framework (CRF)~\cite{Calvary:2002,Calvary:2003}. The use of these two classes aims to implement the generation of the model based user interface. The \textsf{Reification} has two abstract methods. The first one generates all the entire models that we can reify from the \textsf{InputModel} (which extends \textsf{Model}). The second method generates all the partial models that we can reify from the \textsf{InputModel} depending on the place where the user is in the dialog (not shown at this level).

The third abstract class \textsf{AdaptingInterface} is based on the \textsc{Isatine} framework for UI adaptation~\cite{Lopez:2007}. The implementation of this class defines when we trigger an adaptation, how to define the adaptations (\textsf{triggerAdaptation()}), which adaptation is the best one (\textsf{calculateScore()}) and how to evaluate the adaptation (\textsf{triggerEvaluation()}). It also instantiates the Java \textsf{JFrame} object and thus this class is the brain of our framework where different concepts are joined to work together. It is also in \textsf{AdaptingInterface} that we have access to the previous actions executed by the user (the attribute is named \textsf{previousActions}, it is an instance of \textsf{ActionMonitoringList}). 

Now we will go deeper in the implementation. We will start by describing the classes that extend \textsf{Model}.

\subsection*{A.1 Metamodels implemented in \tp}
\label{app:metamodels}

Task model is the highest abstraction level in the Cameleon Reference Framework. Modeling task consists mainly in representing the user goal via a hierarchy of user actions that need to be performed on/with domain objects in a specific temporal and logical order. It is a Platform Independent Model.
Abstract UI: the abstract UI is the result of reification of the task model. It represents an abstract depiction of the UI in terms of Abstract Interaction Units, as well as the relationships among them without consideration for  the context of use. These AIUs are independent of any implementation technology or modality. 
The Final UI (FUI) is the final outcome of the reification process that covers the inference from high-level abstract descriptions to run-time code. The FUI model expresses the UI in terms of implementation technology dependent source code.

\begin{figure*}
	\centering
	\includegraphics[width=.95\textwidth]{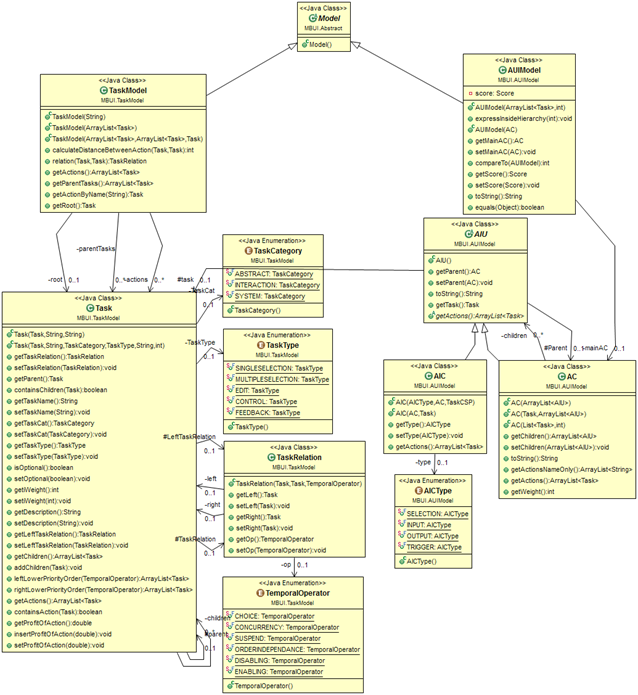}
	\vspace{-8pt}
	\caption{ \tp Task and Abstract UI Metamodels as a subset of \cite{W3C-Task:2014} and \cite{W3C-Abstract:2014}.}
    \label{fig:task-aui}
\end{figure*}

The left part of the schema implements the task model. The right part of the schema implements the abstract user interface (AUI) model. The task model part groups the following classes (\autoref{fig:task-aui}):

\begin{itemize}
\item	\textsf{TaskModel}: extends the “Model” abstract class. It contains a reference to the root “Task” and references to the actions “Task”. It implements also some tools such as the distance in term of nodes between two tasks in the task tree.
\item	\textsf{Task}: this class represents a node or a leaf node in the task model. The “Task” class implements as well the task concept than the action concept. The task is defined by its left and right “TaskRelation”, the “TaskCategory” and the “TaskType”. Each of these classes refer to a concept described in the state of the art chapter. A task is identified by a unique name. There cannot be two tasks with the same name in the task model, this is an assumption that we have made. We also specify the description of the task and the optional nature of the task.
\item	\textsf{TaskCategory}: this class specifies some categories defined in \cite{W3C-Task:2014}.
\item	\textsf{TaskType}: this class specifies some task types defined in \cite{W3C-Task:2014}, we choose to keep only some of them to simplify the problem. Indeed adding all the task types would have increased the number of widget in the FUI and complicated the reification process. 
\item	\textsf{TaskRelation}: this class defines the relation between two tasks having the same parent. It defines the temporal operator that constraints the dialog between the two tasks.
\item	\textsf{TemporalOperator}: this class defines the temporal constraints, they are described by order of priority in the code.
\end{itemize}

The AUI model part groups the following classes (\autoref{fig:task-aui}):
\begin{itemize}
\item	\textsf{AUIModel}: this class extends the “Model” abstract class. It contains a reference to the root container “AC”. It also contains its score and the possibility to compare himself with another “AUIModel” according to this score. The “AUIModel” implements an interesting method called “expressInsideHierarchy()”. This method allows refining recursively an abstract container according to the structure of the task tree.
\item	If we have an AUI model containing an AC [T12 T11 T21 T22]. By calling expressInsideHierarchy(), the AC will be redefined as [[T12 T11][T21 T22]]. This recursive definition of the AC will be used at the FUI level to border the actions and add new data.
\item	\textsf{AIU}: AIU means Abstract Interaction Unit. It is an abstract class that is implemented by AC and AIC. It contains attributes needed by these two classes : a reference to the parent container (AC) and a reference to the task or action (Task) at which this AIU element refers. 
\item	\textsf{AC}: means Abstract Container, it extends AIU. Its only attribute is a list of AIU elements. Thus each element of the list can be either an AC or an AIC. There are two assessors concerning tasks, one called getTask() implemented in AIU, and another getActions() to retrieve the actions (Task) contained in the AIC children of the AC. 
\item	\textsf{AIC}: means Abstract Individual Component, it extends AIU. It contains the type of the AIC (AICType) which defines the semantic meaning of the AIC. This type is determined in his constructor with the action task type who is given in input.
\item	\textsf{AICType}: it defines the four types of AIC described in the state of the art chapter.
\end{itemize}

We hereby hypothesise to bypass the CUI level. We also ignore the dialog reification from the AUI level. Indeed we can notice that unlike the task model with the temporal operator, the AUI model does not expresses the dialog constraints. This is the reason why we save the task associated to each AIU, we can later use the task to infer the dialog.

\begin{figure*}
	\centering
	\includegraphics[width=\textwidth]{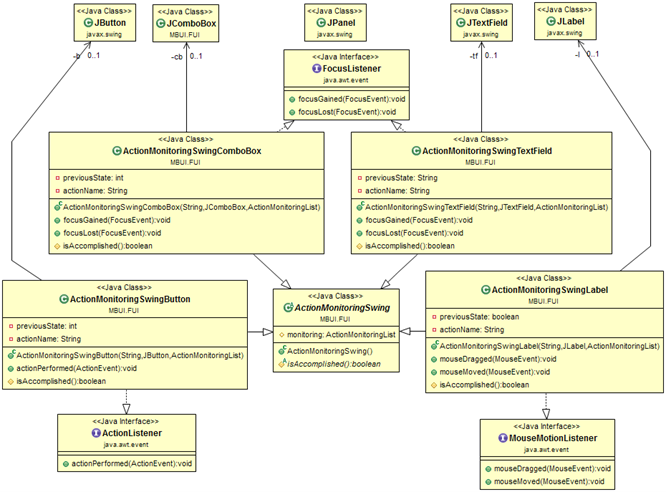}
	\vspace{-10pt}
 \label{fig:fui}
	\caption{\tp Final UI Metamodel as a subset of UsiXML~\cite{Limbourg:2004}.}
\end{figure*}

The Final UI (\autoref{fig:fui}) does not extend the abstract class “Model” because the FUI is implemented by using already existing tools of the development language (such as “JPanel”, “JButton”, “JLabel”, “JComboBox”, “JTextField”). 

\begin{figure*}
    \centering
    \includegraphics[width=\textwidth]{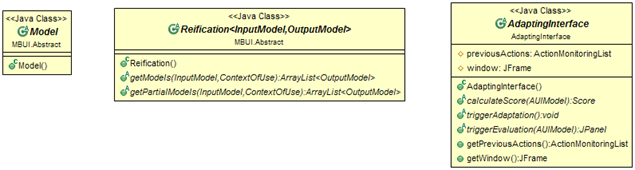}
    \vspace{-8pt}
    \caption{The three top abstract classes implemented in \tp.}
    \label{fig:top}
\end{figure*}
The other classes than the top ones (\autoref{fig:top}) define the listeners to add to each instance of the Swing widgets in order to monitor the user interactions with those widgets and so determining when the action associated to the widget is executed. We can see that each of their constructors takes the “ActionMonitoringList” object in parameter. When the action is considered as executed by the listener, the task name is added to the “ActionMonitoringList”. The five classes are as follows (\autoref{fig:dialog}):

\begin{itemize}
\item \textsf{ActionMonitoringSwing}: this class defines the skeleton of the other widget listener classes. It is an abstract class that contains one reference: the “ActionMonitoringList” that will be used by the implementation of the abstract method “isAccomplished()”.
\item \textsf{ActionMonitoringSwingButton}: this class extends ActionMonitoringSwing. This class listens to the interaction between the user and a button (JButton) and adds the action name to “ActionMonitoringList” when the button is triggered.
\item \textsf{ActionMonitoringSwingLabel}: this class extends ActionMonitoringSwing. This class listens to the mouse movements to monitor if the user has seen the text displayed by the label. When it is considered as read, the action name is added to the “ActionMonitoringList”.
\item \textsf{ActionMonitoringSwingTextField}: this class extends ActionMonitoringSwing. This class uses the focus listener to check if the textbox is filled in. If the textbox is empty then the action name is removed from the “ActionMonitoringList”. If the textbox content remains the same then the listener does nothing. Otherwise, the action name is added to the list.
\begin{figure*}
	\centering
	\includegraphics[width=.99\textwidth]{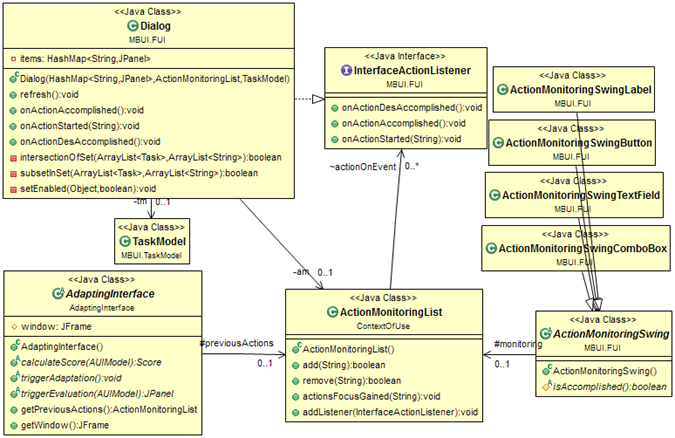}
	\caption{\tp Dialog UI Metamodel.}
        \label{fig:dialog}
\end{figure*}

\item \textsf{ActionMonitoringSwingComboBox}: this class extends ActionMonitoringSwing. The principle is the same than above, if the user does not select a value then the action name is removed. If the user selection is the same than the previous value, nothing is done. In the other case, the action name is added to the list.

\end{itemize}
We will see how we have implemented the dialog of the widgets. We are not talking about the dialog inside the widget itself, but the dialog between the widgets, the consequences on the other widgets of using one widget according to the constraint stated in the task model. This dialog concerns only the widgets because each of them are related to an action and to temporal operators. Consequently the way we deal the dialog here is independent of the containers content and of what is displayed on the screen. The dialog uses the list of monitored actions and the task model as input. Here are the classes involved in the process (\autoref{fig:dialog}):

\begin{itemize}
\item \textsf{ActionMonitoringSwing}: the right part has been already described. This part concerns the user actions monitoring. The abstract class “ActionMonitoringSwing” needs an instance of the “ActionMonitoringList”  to let its child classes adding their action name to the list when the user fulfills it.
\item \textsf{ActionMonitoringList}: this class extends “ArrayList<String>”, a list of strings. Each string contains an action name. We can add or remove an action name of the list according to the user behavior. The class implements also a listener pattern, in order to do that, the class has as attribute a list of listeners called “InterfaceActionListener”. We overwrite the “add()” and “remove()” methods of the ArrayList to call respectively the “onActionAccomplished()” and “onActionDeAccomplished()” methods belonging to each listener contained in the list.
\item	\textsf{InterfaceActionListener}: this interface defines methods that will be implemented by classes who want to be aware of the user actions accomplishment (like “Dialog”). Each method will be called by the “ActionMonitoringList” according to the edition operation triggered on the action list.
\item \textsf{AdaptingInterface}: the “ActionMonitoringList” is an attribute of this class.
\item	\textsf{Dialog}: this class is the heart dealing with the widgets dialog. The class contains three attributes:  the task model, the list of accomplished actions and an hashmap. This hashmap links to each action name a JPanel that contains the widget associated to this action. When an action is accomplished “onActionAccomplished()” or when the “refresh()” method is called, a process starts. This process analyses each temporal operator of the task model and according to its priority, it collects all the left and right actions concerned by this operator. Next, the process compares the set of actions already completed with the right actions and/or the left actions. The deactivation (or activation) of the JPanels (and its content) are followed according to the operator constraint.
\end{itemize}

Example: $T1 ||| T2 >> T3 >> T4$ (all tasks are actions linked to the same root task)
The process analyses the first >> operator.
It collects the left and right actions concerned by the operator according to the operator priority (highlighted by brackets). $ [ T1 ||| T2 ] >> [T3] >> T4$
Next, depending to the actions accomplished by the user (shortcut: \textsf{actionsDone}).

\begin{itemize}
\item	If the left part $[T1 ||| T2]$ is a subset of \textsf{actionsDone} (T1 T2 T3) and an element of \textsf{actionsDone} is in the right part [T3]:
\item	The JPanels associated to the actions of the left part are deactivated.
\item 	The JPanels associated to the actions of the right part are activated.
\item	Else If the left part $[T1 ||| T2]$  is a subset of actionsDone (\eg T1 T2).
\item	The JPanels associated to the actions of the left and the right parts are activated.
\item	Else (\eg \textsf{actionsDone} is empty or T1 or T2).
\item	The JPanels associated to the actions of the left part are activated.
\item	The JPanels associated to the actions of the right part are deactivated.
\end{itemize}

\begin{figure*}
	\centering
	\includegraphics[width=.7\textwidth]{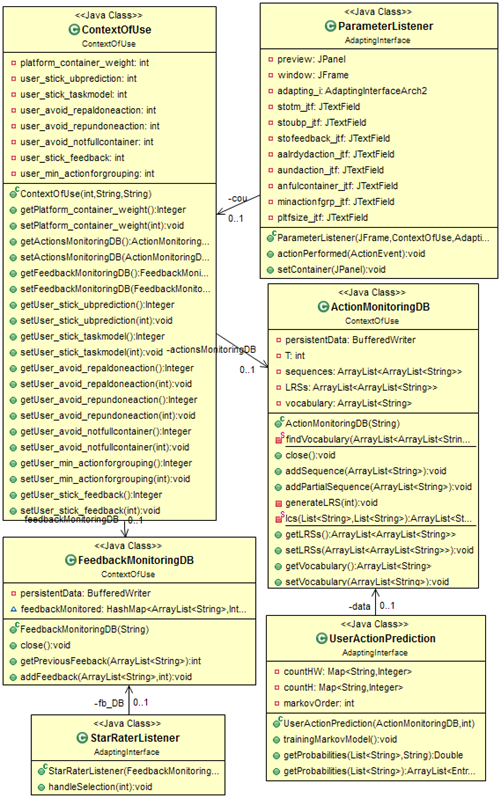}
	\vspace{-8pt}
	\caption{\tp Context Metamodel.}
        \label{fig:context}
\end{figure*}

\begin{figure*}
	\centering
	\includegraphics[width=.85\textwidth]{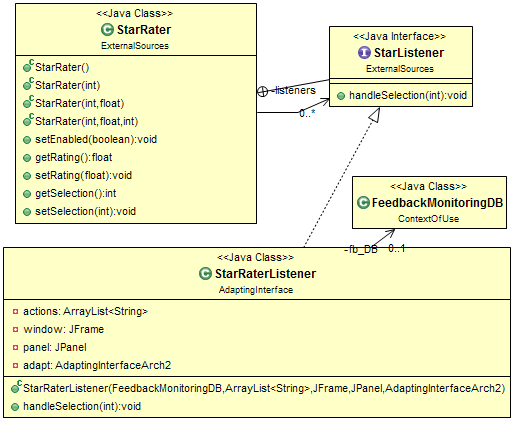}
	\vspace{-8pt}
	\caption{\tp Feedback Metamodel.}
         \label{fig:feedback}
\end{figure*}

\begin{figure*}
	\centering
	\includegraphics[width=.9\textwidth]{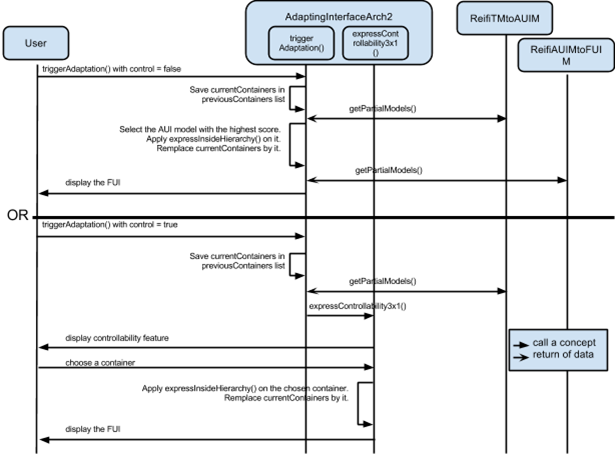}
	\vspace{-10pt}
	\caption{\tp Trigger Adaptation Mechanism.}
        \label{fig:trigger}
\end{figure*}

The metamodels for modeling and storing the context of use, the end user's feedback, and the mechanism to trigger adaptation are represented as UML V2.5 class diagrams and activity diagram in \autoref{fig:context}, \autoref{fig:feedback}, and \autoref{fig:trigger}, respectively.

\end{document}